\documentclass{aa}  
\usepackage{subfig}
\usepackage{subcaption}

\usepackage{graphicx}
\usepackage{multirow}
\usepackage{csvsimple}
\usepackage{url}
\usepackage{wasysym}
\usepackage{amssymb}
\usepackage[hyphenbreaks]{breakurl}
\usepackage{txfonts}
\usepackage{natbib,twoopt}
\usepackage[breaklinks=true]{hyperref} 
\usepackage{csvsimple}
\usepackage{booktabs}

\newcommand{\Vrad}{$V_{\rm Rad}$}
\newcommand{\alfa}{$\alpha$}
\newcommand{\AF}{[\alfa/Fe]}
\newcommand{\Meta}{[M/H]} 
\newcommand{\FeH}{[Fe/H]}
\newcommand{\PbM}{[Pb/Fe]}

\newcommand{\SNR}{$S/N$}

\newcommand{\Gaia}{{\it Gaia}}

\newcommand{\T}{$T_{\rm eff}$}
\newcommand{\g}{log($g$)}

\newcommand{\Vmi}{$V_{\text{micro}}$}

\newcommand{\QC}{$Q_{50}$}
\newcommand{\QS}{$Q_{16}$}
\newcommand{\QQ}{$Q_{84}$}
\newcommand{\SMC}{$\sigma_{MC}$}
\newcommand{\CCF}{$FWHM_{CCF}$}
\newcommand{\D}{$\Delta_{Atm}$}

\newcommand{\CrII}{Cr~{\sc ii}}
\newcommand{\FeII}{Fe~{\sc ii}}
\newcommand{\PbI}{Pb~{\sc i}}
\newcommand{\PbII}{Pb~{\sc ii}}
\newcommand{\FeI}{Fe~{\sc i}}
\newcommand{\VI}{V~{\sc i}}
\newcommand{\CoI}{Co~{\sc i}}

\bibpunct{(}{)}{;}{a}{}{,} 
\makeatletter
\newcommandtwoopt{\citeads}[3][][]{\href{http://adsabs.harvard.edu/abs/#3}%
        {\def\hyper@linkstart##1##2{}%
                \let\hyper@linkend\@empty\citealp[#1][#2]{#3}}}
\newcommandtwoopt{\citepads}[3][][]{\href{http://adsabs.harvard.edu/abs/#3}%
        {\def\hyper@linkstart##1##2{}%
                \let\hyper@linkend\@empty\citep[#1][#2]{#3}}}
\newcommandtwoopt{\citetads}[3][][]{\href{http://adsabs.harvard.edu/abs/#3}%
        {\def\hyper@linkstart##1##2{}%
                \let\hyper@linkend\@empty\citet[#1][#2]{#3}}}
\newcommandtwoopt{\citeyearads}[3][][]%
{\href{http://adsabs.harvard.edu/abs/#3}
        {\def\hyper@linkstart##1##2{}%
                \let\hyper@linkend\@empty\citeyear[#1][#2]{#3}}}
\makeatother

\begin{document}

   \title{The AMBRE Project: Lead abundance in Galactic stars}


   \author{G. Contursi\inst{1}\thanks{Send offprint requests to Patrick de Laverny (laverny@oca.eu)}
                        \and
          P. de Laverny\inst{1}
          \and 
          A. Recio-Blanco\inst{1}
          \and
          M. Molero\inst{2,3}
         \and
          E. Spitoni\inst{3}     
          \and
          F. Matteucci\inst{3,4,5}    
           \and
          S. Cristallo\inst{6,7}           
          }

   \institute{
   Université Côte d'Azur, Observatoire de la Côte d'Azur, CNRS, Laboratoire Lagrange, Bd de l'Observatoire, CS 34229, 06304 Nice cedex 4, France
         \and
         Institut f\"ur Kernphysik, Technische Universit\"at Darmstadt, Schlossgartenstr. 2, Darmstadt 64289, Germany
         \and I.N.A.F. Osservatorio Astronomico di Trieste, via G.B. Tiepolo  11, 34131, Trieste, Italy \and Dipartimento di Fisica, Sezione di Astronomia,
  Universit\`a di Trieste, Via G.~B. Tiepolo 11, I-34143 Trieste,
  Italy
\and I.N.F.N. Sezione di Trieste, via Valerio 2, 34134 Trieste, Italy
\and I.N.A.F. Osservatorio Astronomico d'Abruzzo, via Maggini snc, 64100 Teramo, Italy
 \and I.N.F.N. Sezione di Perugia, via Pascoli snc, 06123 Perugia, Italy
            }

   \date{Received ??; accepted ??}

 
  \abstract
   {The chemical evolution of neutron capture elements in the Milky Way  is still a matter of debate. Although more and more studies investigate their chemical behaviour, there is still a lack of a significant large sample of abundances of a key heavy element: lead. }
   {Lead is the final product of the s-process nucleosynthesis channel and is one of the most stable heavy elements. The goal of this article is to present the largest catalogue of homogeneous Pb abundances, in particular for metallicities higher than -1.0~dex, and then to study the lead content of the Milky Way.}
   {We analysed high-resolution spectra from the ESO UVES and FEROS archives. Atmospheric parameters were taken from the AMBRE parametrisation. We used the automated abundance method GAUGUIN to derive lead abundances in 653 slow-rotating FGK-type stars from the 368.34~nm Pb I line.}
   {We present the largest catalogue ($\sim$650 stars) of Local Thermodynamic Equilibrium (LTE) and non-LTE lead abundances ever published with metallicities ranging from -2.9 to 0.6 dex and [Pb/Fe] from -0.7 to 3.3 dex. Within this sample, no lead-enhanced Asymptotic Giant Branch (AGB) stars were found, but nine lead-enhanced metal-poor stars ([Pb/Fe] > 1.5) were detected. Most of them were already identified as carbon-enhanced metal-poor stars with enrichments in other $s$-process species. The lead abundance of 13 \Gaia\ Benchmark Stars are also provided. We then investigated the Pb content of the Milky Way disc by computing vertical and radial gradients and found a slightly decreasing [Pb/Fe] radial trend with metallicity. This trend together with other related ratios ([Pb/Eu], [Pb/Ba], and 
   [Pb/$\alpha$]) are interpreted thanks to chemical evolution models. The two-infall model closely reproduces the observed trends with respect to the metallicity. It is also found that the AGB contribution to the Pb Galactic enrichment has to be strongly reduced. Moreover, the contribution of massive stars with rather high rotational velocities should be favoured in the low-metallicity regime.}
  {}


   \maketitle
%

\section{Introduction}
\label{sec:Introduction}

The nucleosynthesis of elements heavier than iron cannot occur by charged particle reactions;
it is mainly carried out via neutron capture. According to the seminal work of \citet{Burbidge57}, neutron capture occurs through two processes: the slow ($s$-) process and the rapid ($r$-) process. The latter takes place at high neutron densities and short  timescales (compared to $\beta$ decay). Several neutrons are added before multiple $\alpha$ and $\beta^-$ decays allowing the isotope to find stability. 

The different sites of $r$-element production are still not clearly identified. Among the different scenarios, one can find mergers of compact objects (neutron stars and/or neutron star--black hole; e.g. \citealt{1999ApJ...525L.121F, Surman08, Thielemann2017}) as well as winds from type II Supernovae (SNIIe; \citealt{Woosley94}) and rare and energetic magneto-rotational SNe (MR-SNe; \citealp{winteler2012, reichert2023}).

On the contrary, the $s$-process occurs on longer timescales and with lower neutron densities. The main production sites for these $s$-elements, in particular the late evolutionary stages of low- and intermediate-mass stars, are better known, although their actual production rates are still debated.\\

The abundance distribution of elements formed by the $s$-process as a function of atomic number shows three peaks: one around Sr-Y-Zr, 
one around Ba-La-Ce, 
and one around Pb-Bi. 
The isotopes of lead and bismuth are the most stable ones formed by the $s$-process. For instance, one of the most stable Pb isotopes is $^{208}$Pb (Z = 82, N\footnote{Neutron number} = 126) because the neutron and proton shells are closed, reducing the cross-section to the next neutron capture.

In this context, each $s$-process peak is associated with a different source. The main sources are rotating massive stars and Asymptotic Giant Branch (AGB) stars. 
First, elements with an atomic mass number (A) between about 60 and 90 are thought to be formed through the $weak$-$s$ process during the He-core burning phase of massive stars (M $\gtrsim$ 8M$_{\astrosun}$). This happens through the $^{22}$Ne($\alpha$,n)$^{25}$Mg reaction [Cameron 1960, Iben and Renzini 1983]. When the helium runs out, the $^{22}$Ne is only partially consumed and neutron production continues in the convective carbon layer \citep{Couch74, Kappeler89, Pignatari10, Kappeler11}. This $^{22}$Ne comes from the $^{14}$N produced during the hydrogen combustion phase in the core via the CNO cycle.  We note that these $s$-process elements are not altered by the final explosive nucleosynthesis of SNIIe and enrich the interstellar medium (ISM) (e.g. \citealt{Kappeler89, Prantzos90, Limongi2018}). 

The $main$-$s$ component is responsible for the production of heavier elements with 90 $\lesssim$ A $\lesssim$ 204 during the thermal pulse phases of low- and intermediate-mass stars in the  last stage of their evolution (AGB with 1.3 M$_{\astrosun}$ $\la$ M $\la$ 8$_{\astrosun}$) \citep{Arlandini99, Busso99}. The protons left by the receding of the convective envelope are captured by the $^{12}$C present in the He-intershell between two thermal pulses (the interpulse). This produces $^{13}$C via the reaction $^{12}$C(p,$\gamma$) $^{13}$N($\beta^+$, $\nu$) $^{13}$C and forms a thin pocket of $^{13}$C. The latter is itself converted to oxygen via the $^{13}$C($\alpha$, n) $^{16}$O reaction, thus producing a source of neutrons for the $s$-process \citet{Straniero95}. 
Since the $^{13}$C pocket is partially covered by a layer of $^{14}$N and $^{22}$Ne ($^{14}$N acts as a poison for neutrons due to the $^{14}$N(n,p) $^{14}$C reaction), the $s$-process takes place in the deepest part of this $^{13}$C pocket where little $^{14}$N is found \citep{Cristallo2009}. 
For intermediate-mass stars (M $\ga$ 3M$_\odot$), the temperature at the base of the convective zone is higher and a second neutron source is activated: the $^{22}$Ne($\alpha$,n)$^{25}$Mg reaction. This source is marginal in low-mass AGBs of solar metallicity. It is important to note that stellar metallicity and rotation have a large impact on the efficiency of these nucleosynthesis channels \citep{Cristallo2009, Cristallo2011, Karakas14}. 

However, neither the weak nor the main $s$-process is sufficient to explain the observed lead abundances (Pb, Z\footnote{Proton number} = 82) in the Solar System. Therefore, a new process (called the strong $s$-process) was introduced \citep{Kappeler89}, in line with the proposition of \citet{CR67} that the Solar System lead is not formed by the $r$-process. Later studies on AGB nucleosynthesis and yields indeed confirmed that Pb production in low-mass and low-metallicity stars corresponds to this strong $s$-process \citep{Gallino98, 2000A&A...362..599G, Busso01, Choplin22}, whose source is still not well identified.\\


\indent 
Furthermore, observational studies of lead abundances in stellar atmospheres are not very common, mainly because of the difficulty  obtaining reliable results, due to the complexity of the spectral domain in which the main (and weak) atomic transitions of Pb reside. 
We note that most of the studies related below  deal with
metal-poor stars enriched in carbon (known as CEMP stars). These stars are probably affected by mass transfer from an AGB companion that  previously produced carbon and neutron-capture elements. Among the main Pb studies, one can cite the work of \citet{VanEck01} who analysed three metal-poor stars (HD187861, HD196944, and HD224959) with very high Pb abundances. 
This study was completed by the analysis of two other Pb-rich stars \citep{VanEck03}. Independently, \citet{Jonsell06}, \citet{Lucatello03}, \citet{Sivarani04}, \citet{Ivans05}, and \citet{Barbuy11} added a few new Pb-rich stars to this list. 
Most of these stars are very metal-poor ([M/H] < -2.0 dex), highlighting again the  $s$-process contribution of low-metallicity environments. 

Independently, \citet{Aoki02} studied eight metal-poor (-2.7 < [Fe/H]< -1.9 dex) carbon-rich stars and found a strong dispersion of [Pb/Ba]. \citet{Allen06} studied 26 Ba-type stars, enriched in $r$-elements and Pb. \citet{Roederer09} determined Pb (and other heavy element) abundances for 27 metal-poor stars (-3.1 < [Fe/H] < -1.4 dex) with heavy element enrichments. Finally, \citet{Roederer14}   provided a catalogue of 13 metal-poor stars for which atmospheric parameters and abundances (including some Pb abundances) are provided.  
Non Local Thermodynamic Equilibrium (NLTE) corrections for some Pb lines were also   provided by \citet{Mashonkina12}. More recently, \citet{Roederer20} and \citet{Peterson21} determined Pb abundances for three and four stars, respectively, based on the analysis of UV lines. 
{Finally, we recall that \cite{DeSmedt15, DeSmedt16}   also reported Pb abundances in some post-AGB stars.
Lastly, we recall that \citet{Prantzos18} published Galactic chemical evolution models of \PbM\ versus \FeH.  As  can be seen  in their Fig.~16, there is a clear paucity of observational data, especially at metallicities higher than -1.0.\\

In this work we provide Pb abundances within the context of the AMBRE Project \citep{AMBRE13}. AMBRE has  parametrised ESO archived stellar spectra (FEROS, UVES, and HARPS), thanks to the MATISSE algorithm \citep{ARB06} trained with a specific synthetic spectrum grid \citep{PDL12}. Within AMBRE, chemical abundances are derived using the GAUGUIN optimisation method \citep[see Sect.~\ref{GAUGUIN} and][for a detailed description]{2012ada..confE...2B, RB16} as already performed, for instance, for the determination of Li, $r$-process, Mg, and S chemical abundances \citep[respectively]{Guiglion16, Guiglion18, Pablo20, Perdigon21}.\footnote{However we recall that AMBRE iron-peak element abundances were derived thanks to a more classical technic based on line profile fitting that was partially automated \citep{Sarunas17}.} The initial goal of AMBRE was to test both algorithms, MATISSE and GAUGUIN, on large spectra datasets and to develop an optimal stellar parametrisation pipeline for the analysis of \Gaia\ Radial Velocity Spectrometer data by the DPAC/GSP-spec module \citep{GSPspecDR3}.
The adopted methodology for lead abundance estimations is presented in Sect.~\ref{Sect:Derivation}, while Sect.~\ref{Sect:Cat} describes the AMBRE:Pb catalogue. Section~\ref{Sect:Pec} discusses  Pb abundances in some peculiar stars, including the \Gaia\ Benchmark Stars, and Sect.~\ref{Sect:MW} explores the lead abundance properties in the Milky Way. Our conclusions are summarised in Sect.~\ref{Sect:Ccl}.

\section{Derivation of lead abundances} \label{Sect:Derivation}

For this study, we adopted the classical AMBRE methodology, basically the abundance determination thanks to the GAUGUIN method and a specific synthetic reference spectra grid with varying Pb abundances, for the determination of lead abundances.

Several lead lines were previously analysed in stellar spectra. For instance, in
the ultraviolet domain (not covered by the AMBRE spectra), the \PbII~220.35~nm was adopted by \citet{Roederer20} in three metal-poor stars, whereas the \PbI~283.305~nm was analysed by \citet{Roederer09}, \citet{Roederer16} and \citet{Peterson21} in 27, one and four metal-poor stars, respectively.  Other lead lines in the optical domain such as the 363.95, 373.99, 405.78 and 722.89~nm transitions were also analysed to derive the Pb solar abundance \citep[for instance]{DeJager62, Grevesse69, HS73}. Nevertheless, all those atomic transitions are severely blended or very faint in the solar-type stellar spectra. To avoid such problems, we have therefore selected the \PbI~line at 368.346~nm, one of the less blended lines in cool star (particularly dwarfs) spectra and the most accurate one for deriving lead abundance in the Sun (e.g. \citet{Asplund09, Grevesse15, Asplund21}). For this line, we adopted the atomic line data of \citet[][see Sect.~\ref{Grid} for more details]{Biemont00}. Some example observed spectra around the selected \PbI\ line are shown in Fig. \ref{fig:FigPb}.
\begin{figure}[t]
        \centering
        \includegraphics[scale = 0.35]{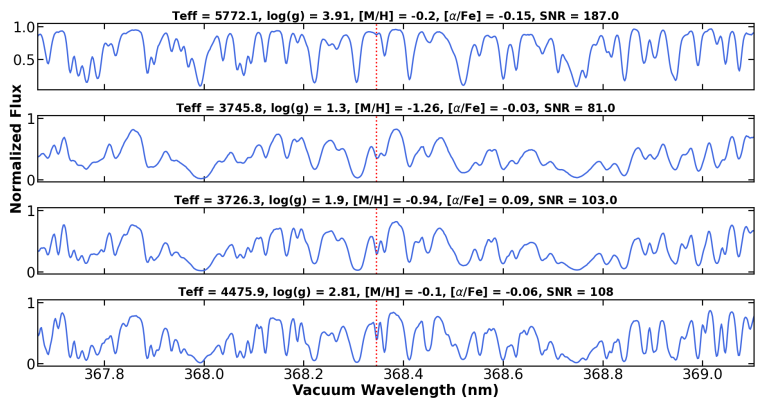}
        \caption{Identification of the selected \PbI\ line (red vertical line) in some AMBRE example spectra.}
        \label{fig:FigPb}
\end{figure}


\subsection{Lead abundances determination with GAUGUIN \label{GAUGUIN}}
        
Briefly, to derive chemical abundances, GAUGUIN first builds a set of reference spectra with varying lead abundances
for the atmospheric parameter of the analysed star (hereafter called the 1D grid). These parameters (\T, \g, \Meta\footnote{Within the AMBRE parametrisation: \Meta=[Fe/H].} and \AF) are provided by the AMBRE parametrisation.  The reference spectra of the 1D grid are interpolated from a 5D grid (four atmospheric parameters plus \PbM) of synthetic spectra, described hereafter.
Then, the observed spectra are automatically normalised over $\sim$4~nm and a second normalisation is performed over a smaller range around the line ($\sim$ 0.06 nm in our case) to locally readjust the continuum \citep{Pablo20}. 
Finally, the minimum of the quadratic distances between the observed and reference spectra is calculated over a spectral range of  0.02~nm around the lead line. This provides the initial guess of the lead abundance, from which the Gauss-Newton algorithm derives the final solution
in a similar way as 
implemented for the \Gaia/GSP-spec pipeline \citep{GSPspecDR3}.

Together with the chemical abundances, GAUGUIN also provides a value for the upper limit of the abundance below which the line cannot be detected. This upper limit is mainly dependent on the atmospheric parameters and the signal-to-noise ratio (\SNR) which is also estimated by the AMBRE parametrisation. It is used to reject weak and/or doubtful detections.
Moreover, for each spectrum, a white Gaussian noise (whose standard deviation is proportional to the \SNR~of the input spectrum) was computed for 1,000 realisations and added to the initial spectrum. GAUGUIN then determined the lead abundance for each realisation, leading to 1,000 abundances for each spectrum. We then calculated the median value (represented by the quantile 50, \QC, and adopted as the lead abundance hereafter) as well as the quantiles \QQ~and \QS~which can be used to estimate the error associated with the determination of lead abundances (see Sect.~\ref{Error}). 

\subsection{Observed spectra \label{Obs}}

As described above, our method compares observed and synthetic reference spectra. Both sets of spectra have thus to be adapted in order to have similar spectral resolution and wavelength sampling.

First, we have selected AMBRE spectra covering the selected \PbI\ line. Our sample of spectra were
collected with the FEROS and UVES spectrographs (BLUE346 and BLUE390 setups)
already parametrised within the AMBRE Project (\citet{FEROS} and \citet{UVES} respectively). 
We first selected spectra having a \SNR $>$ 20, 3500~K $<$ \T~$<$ 8000~K, 0.0 $<$ \g~$<$ 5.5 ($g$ in cm/s$^2$) and whose AMBRE Quality Flag is 0 or 1 (i.e. good or very good parametrisation; following the AMBRE flagging criteria). 
This led to an initial sample of 5,998 FEROS spectra plus 1,508 and 1,202 UVES spectra of the BLUE346 and BLUE390 setups, respectively. GAUGUIN thus analysed a total of 8,708 stellar spectra times the number of Monte-Carlo \SNR\ realisations (about 8.7 million spectra in total). We note that the corresponding number of stars is much smaller because of the presence of several repeated spectra for several stars (see below).



To be able to perform the chemical analysis, the observed spectra were first corrected for radial velocity shifts (whose value is also provided by the AMBRE parametrisation) and a spectral domain of 4~nm around the lead line was selected. As the spectral resolutions of the FEROS and UVES spectra slightly differ (R$\sim$ 48,000 and R$\sim$ 40,000, respectively), the FEROS spectra were then degraded to the smallest resolution in common (i.e. the UVES resolution).
We also adopted a sampling with a wavelength step of 0.02~nm in order to fulfil the Nyquist-Shannon criterion.

\subsection{Reference synthetic grid and adopted linelist}
\label{Grid}
Independently, we computed a high-spectral resolution reference grid of synthetic spectra with an initial wavelength step of 0.001~nm and covering 4~nm around our selected lead line.
This grid was convolved to mimic a spectral resolution of 40,000 and resampled as the observed spectra.  The covered atmospheric parameter ranges are 3600K $\le$ \T~$\le$ 8000K (in steps of 250K if \T~> 4000K, 200~K below), +0.0 $\le$ \g~$\le$ +5.5 (in steps of 0.5), and - 2.5 $\le$ \Meta~$\le$ +1.0 dex (with steps of 0.25 dex). Up to 13 values of \AF~were considered for each value of the metallicity depending on the availability of the MARCS models (with steps of 0.1 dex). For each combination of these main atmospheric parameters, we computed 21 spectra with lead abundances varying between -2.0 $\le$ \PbM~$\le$ 2.0 with a step of 0.2~dex. The AMBRE:Pb 5D grid is composed of 478,400 synthetic reference spectra.\\
 
For this grid computation, we adopted the version 19.1.4 of TURBOSPECTRUM \citep{TURBO} which computes continuum opacity and solves the radiative transfer equation in the lines for a given atmospheric model assuming hydrostatic equilibrium and Local Thermodynamic Equilibrium (LTE). We also adopted 1D MARCS models \citep{MARCS} computed under two geometries: plane-parallel and spherical (\g~$\ge$ 3.5 and \g~< 3.5, respectively). 
We adopted solar abundances and isotopic compositions from \citet{Grevesse07} and a micro-turbulence velocity relation that depends on \T, \g~and \Meta~using the second version of the GES empirical relation based on micro-turbulence velocity determinations from literature samples (M. Bergemann, private communication).

Moreover, a specific atomic and molecular line list has been built for the computation of this spectra grid. As the AMBRE stellar sample covers a wide range of atmospheric parameters, a careful examination of atomic and molecular contributions to the emerging spectrum was performed.
We first started with the atomic line list extracted from the Vienna Atomic Line Database (VALD\footnote{\url{http://vald.astro.uu.se/}}; \citealt{VALD95, VALD00, VALD15}). 
For the \PbI~ line, we adopted the atomic data of \citet{Biemont00} and the hyperfine structure (hfs) as well as the isotope shift from \citet{Manning50}. A rather similar lead line list was, for instance, used by \citet{Jonsell06}. For the molecular transitions, we adopted the line lists from the following species: $^{12}$C$^{12}$C, $^{12}$C$^{13}$C,$^{13}$C$^{13}$C \citep{Brooke13}, $^{12}$C$^{14}$N, $^{12}$C$^{15}$N, $^{13}$C$^{14}$N \citep{Sneden14}, $^{12}$CH, $^{13}$CH \citep{Masseron14}, $^{16}$OH (Masseron, priv. comm), AlH, SiH \citep{Kurucz92}, VO \citep{VO2016} and TiO (\citet{TiO}, including its 46-50 isotopologues).

The selected lead line is blended by an iron line in its blue wing (around 368.362~nm) and, in its red wing around 368.306 nm, by contributions from \CoI, \FeI\ and \VI\ lines.  To improve possible mismatches between observed and synthetic
spectra, we astrophysically calibrated the oscillator strength of 13 main atomic contributions over 0.5~nm around the Pb line using solar and Arcturus spectra.  
No astrophysical calibration of the lead line atomic data were performed because of their high-quality.
The solar observed spectrum is the \citet{Wallace11} one whereas we adopted the S4N's Arcturus spectrum \citet{S4N2004}. Both spectra were degraded to R = 100,000 and 40,000 for the calibration. Synthetic spectra have been computed adopting the atmospheric parameters and chemical abundances reported in Table 1 of \citep{BestArticleEver}. Table \ref{ModifLines} presents the lines that were selected for calibration as well as the atomic data adopted for the lead transitions.

\begin{table}[t!]
        \caption{\label{ModifLines} Line data for the atomic features around the \PbI\ line that were calibrated.}
        \centering
        \begin{tabular}{lccc}
                \hline
                \hline
                Element & Air Wavelength & \multicolumn{1}{c}{\it E} & \multicolumn{1}{c}{Adopted \it loggf}\\
                 & (nm) & (ev) & (dex)  \\
                \hline
                \it Calib. lines & & &\\
                \CrII & 368.4224  & 4.944 & -1.200 \\
                \FeI  & 368.2167  & 2.998 & -1.200 \\ 
                \FeI  & 368.2210  & 2.940 & -0.200 \\
                \FeI  & 368.2212  & 3.301 & -2.400 \\
                \FeI  & 368.2244  & 3.546 & -0.000 \\
                \FeI  & 368.2738  & 3.301 & -3.300 \\ 
                \FeI  & 368.3054  & 0.052 & -2.700 \\
                \FeI  & 368.3582  & 3.960 & -2.600 \\ 
                \FeI  & 368.3611  & 3.301 & -2.300 \\
                \FeI  & 368.3629  & 2.484 & -2.800 \\  
                \FeI  & 368.4107  & 2.727 & -0.500 \\
                \FeI  & 368.4137  & 3.301 & -0.900 \\
                \FeII & 368.2658  & 4.479 & -2.250 \\
                \it Lead lines & & & \\
                \PbI  & 368.3479  & 0.970 & -2.342 \\
                \PbI  & 368.3470  & 0.970 & -1.239 \\
                \PbI  & 368.3481  & 0.970 & -1.683 \\
                \PbI  & 368.3442  & 0.970 & -1.382 \\
                \PbI  & 368.3460  & 0.970 & -0.754 \\
                \hline
        \end{tabular}
    \tablefoot{Lead atomic data are from \cite{Biemont00} and were not calibrated.}
\end{table}



We also convolved the 5D grid with a rotational broadening profile assuming a rotational velocity of 2.0~km.s$^{-1}$ for each spectrum. Regarding the macroturbulent velocity, we adopted the relation provided by
\citet[][Eq.~8, validity domain: 5,250 to 6,400~K]{Doyle14}
for any dwarf spectra with \g~ larger than 3.7 and also for spectra
having \g~<~4.0 when \T~> 5250K. For the stars hotter than 6,400~K,
we adopted the macroturbulent velocity estimated for this limiting \T \ to avoid any extrapolation.
For giant stars with \g < 3.5, we used Eq.~2 of \citet{HM07} assuming that the luminosity classes II, III and IV correspond respectively to \g~= 1.5, 2.5 and 3.5. We note that this relation is valid as long as the effective temperature is between 4,000~K and 5,100~K. We also used this relation for giants with 3.5 < \g~< 3.7 and \T~< 5250K.
Finally, we adopted a Gaussian macro-turbulent profile and, according to \citet{Takeda17}, we multiply each obtained macroturbulent velocity by a factor 0.6 since the used relations assume radial-tangent broadening profiles.\\

The detectability of the selected lead line in stellar spectra having a spectral resolution of 40,000 is illustrated in Fig. \ref{fig:Pb-detect}. This figure  presents a Kiel diagram colour-coded with the minimum lead abundance (in dex) that could be measured for metallicities varying between -1.0 to 0.0 dex. This threshold was estimated from the grid of synthetic spectra in which we searched for the lead abundance corresponding to a (normalised) flux decrease of 0.5\% at the Pb line core with respect to a reference spectrum with [Pb/Fe] = -2.0 dex (i.e. no Pb and lowest lead abundance in the grid of reference spectra). It can be seen that this lead line seems to be easily detectable ([Pb/Fe] > 0 dex) in stars whose effective temperature is cooler than about 7000K, whatever their surface gravity is.

\begin{figure}[t]
        \centering
        \includegraphics[scale = 0.15]{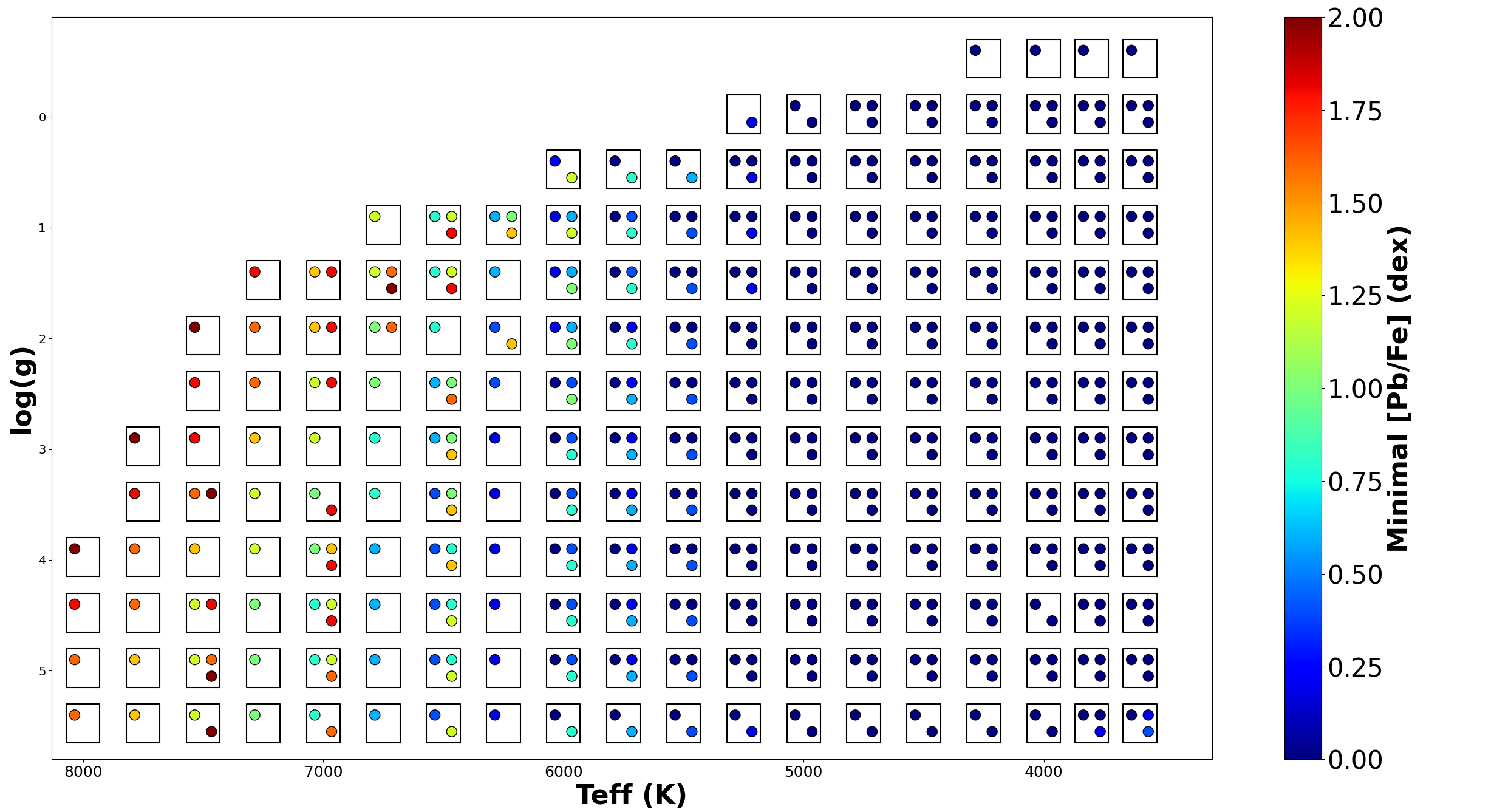}
        \caption{Kiel diagram colour-coded with the lowest lead abundance (in dex) that could be detected in a spectrum whose (normalised) Pb line core flux is 0.5\% deeper than that of a reference spectrum with [Pb/Fe] = -2.0~dex (i.e. negligible amount of Pb, and therefore undetectable line). For each combination of effective temperature and surface gravity, we estimated this lowest Pb abundance for three values of [M/H]: 0.0, -0.5, and -1.0 dex (from top to bottom and left to right in each small square).}
        \label{fig:Pb-detect}
\end{figure}

\subsection{Lead abundance for stars with parameters outside the reference grid}
\label{Sect:ExtremeStars}
During the analysis, about 20 spectra with atmospheric parameters and/or Pb abundances outside the 5D grid borders
were identified (\PbM~$>$ 2.0~dex and/or \FeH~$<$ -2.5~dex). For analysing these specific cases with extreme parameters, we derive their lead abundances in a classical way,
without using GAUGUIN. For that purpose, we computed synthetic spectra with the same tools as described
above. MARCS models were interpolated at the AMBRE atmospheric parameters and synthetic spectra were computed
with various lead abundances to found the best agreement between the normalised observed (provided by GAUGUIN)
 and simulated spectra. This procedure was checked and validated with some stars whose atmospheric parameters are within the grid and for which Pb abundances 
derived with GAUGUIN were available. Both Pb abundance estimates were found to be in perfect agreement, allowing to include these extreme parameter stars within the final catalogue of lead abundances, described below.

\section{The AMBRE lead abundances catalogue}
\label{Sect:Cat}
We present in this section the different steps that have been adopted to construct the AMBRE catalogue of lead abundances (i.e. the selection of the best derived values), the treatment of the repeats (as several spectra are available for some specific stars), the derivation of the uncertainties and the validation by comparison with literature values.

\subsection{Solar lead abundance \label{Sun_Pb}}

Our automatic procedure leads to the fit between the solar observed and synthetic spectra (after calibration of the line list) shown in Fig.~\ref{fig:SunCalib}. We recall that the synthetic grid has been computed without calibrating the lead line atomic data and assuming the solar chemical abundances of \citet{Grevesse07}: {A(Pb)\footnote{A(Pb) = \PbM + A(Pb)$_{\astrosun}$, where A(Pb)$_{\astrosun}$ = $\log$(n$_{Pb}$/n$_{H}$) + 12 with n$_{Pb}$ being the number of lead atoms per unit volume.}=2.00 dex.} We also recall that we analysed the solar spectra of \citet{Wallace11} degraded at R = 40,000. Our derived solar LTE abundance is \PbM~= -0.20~dex, hence, A(Pb) = 1.80. The associated uncertainty is around $\pm$0.01~dex.

This abundance is fully compatible with several previously published solar Pb photospheric values that analysed the same Pb line as we used: \cite[][1D model: A(Pb)=1.85]{AG89}, \cite[][3D model: A(Pb)=1.75]{Asplund09} , \cite[][3D LTE models: A(Pb)=1.80]{Grevesse15}. Nevertheless, we note that solar photospheric abundances are often smaller than meteoritic values obtained from CI carbonaceous chondrites \citep{Lodders09} with differences reaching about 0.25 dex. The rather complex spectral region upon which the lead line stands could explain the observed mismatches in the abundances, such as other physical assumptions as the selected atmospheric models and/or 1D/3D structure, LTE/NLTE effects. For instance, \citet{Mashonkina12} estimated a NLTE correction of up to 0.20 dex for the Sun when adopting the same Pb line as we used.

As a conclusion, our automatic procedure derives a solar photospheric lead abundance that is fully consistent with literature values within less than $\sim$0.05~dex, depending on the adopted physical assumptions. Therefore, no need of any calibration of our Pb abundance was found necessary.
\begin{figure}[t]
        \centering
        \includegraphics[scale = 0.19]{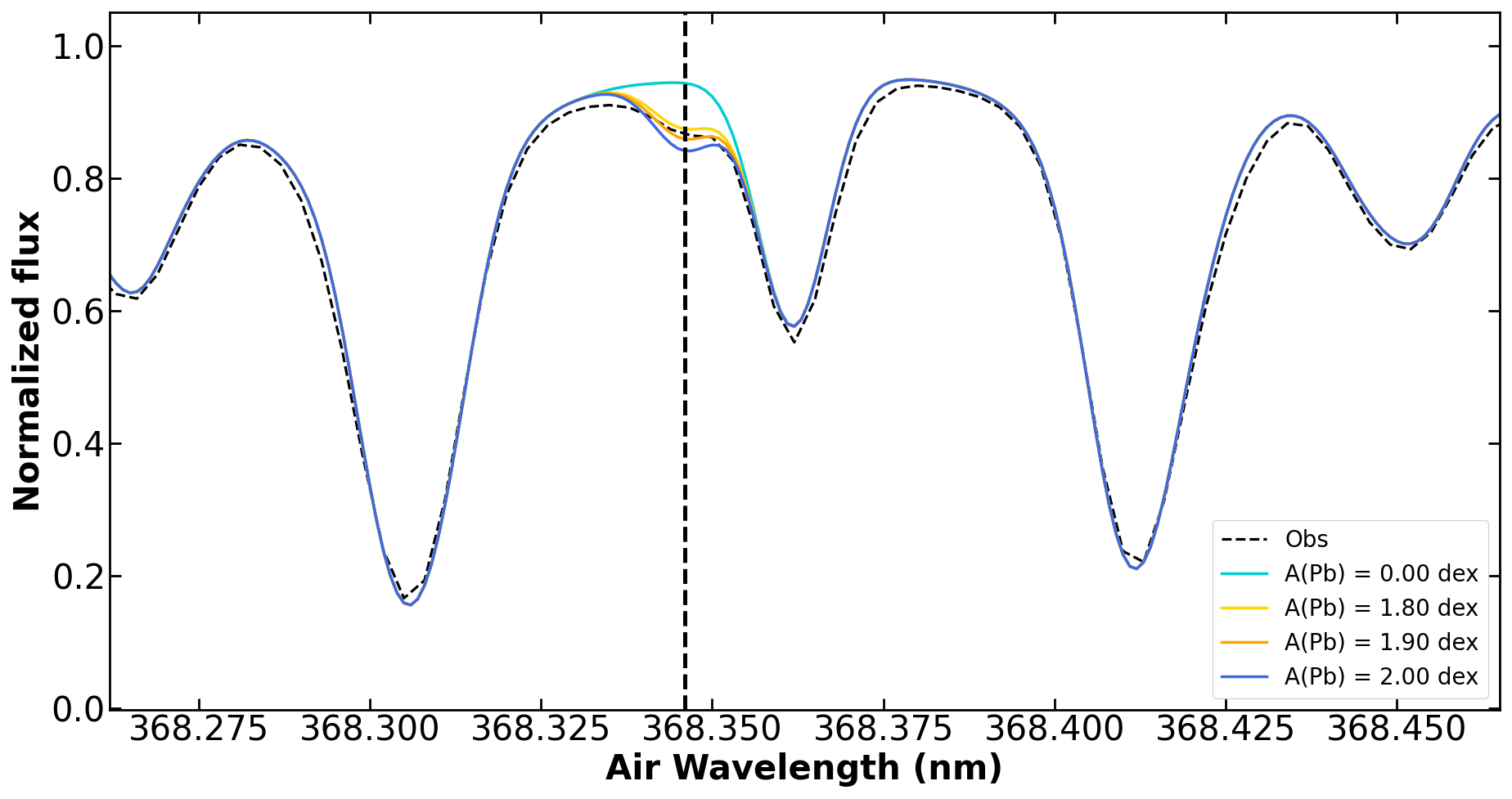}
        \caption{Comparison between the solar observed spectrum from \citet{Wallace11} (black dotted line) and a synthetic spectrum (interpolated to the Sun's atmospheric parameters from \citealt{BestArticleEver}) for different values of A(Pb) = 0.00, 1.80, 1.90, and 2.00.}
        \label{fig:SunCalib}
\end{figure}

\subsection{Selection of the best analysed spectra \label{SelectSpec}}
In some cases, GAUGUIN was unable to determine reliable lead abundances because of too low \SNR\ spectra, too weak lead lines, cosmic rays and/or limitations in the parameters of some specific spectra with respect to the reference grid borders (see Sect.~\ref{Sect:ExtremeStars}). Moreover, a weak lead line can be undetectable in high rotating stars, for which line blending could be extreme in the spectra.

In the following, we thus only considered slow-rotating stars when building the catalogue by filtering with the \CCF~parameter (Full Width at Half Maximum of the Cross-Correlation Function). This quantity is determined during the AMBRE parametrisation when estimating the radial velocity (\Vrad). It takes into account several broadening effects such as rotation and macroturbulent velocities. We therefore selected spectra with \CCF$<$15~km.s$^{-1}$, following previous AMBRE recommendations \citep{Guiglion16, Perdigon21}. This cut removes 13.4\% and 18.4\% spectra of the initial sample of UVES (considering the two setups) and FEROS, respectively (i.e. a total of 1467 spectra are then rejected because of their too large line-broadening).

On the other hand, 
we have systematically rejected any spectra whose abundance value is lower than  2.3 times the abundance upper limit for giants, defined as \g~$\le$ 3.5, and 2.8 times for dwarfs with \g~> 3.5, (see Fig.~\ref{fig:Pb-detect}).
{To select those thresholds, we looked at several observed spectra of various spectral types and visually chose the value above which the line can not be mistaken with noise. In addition, we note that this upper limit also corresponds to the lowest Pb abundance that could be determined for a set of atmospheric parameters and a given S/N, as already discussed in Sect.~\ref{Grid}. 
These rejection criterion were adopted in order to select only stars with the best detected Pb-line, compared to a spectrum without any Pb line detection (abundance upper limit). Since the Pb abundance measurement is fully automatic, some low-quality and/or too faint Pb-line spectra that could have suspicious Pb abundances are then automatically rejected. Such a procedure could reject good measurements  for low lead abundance stars, but it guarantees a high-quality and a clean final sample of Pb abundances. A similar procedure was adopted for the chemical analysis of the \Gaia/RVS spectra \cite[see][for more details]{GSPspecDR3}. 
Such a strict rejection of the faintest Pb signatures led to a final selection of 1341 stellar spectra (15.4\% of the initial sample).

\subsection{Uncertainties of the derived lead abundances \label{Error}}
Various sources of uncertainties could impact our abundance determination
and their respective importance is evaluated hereafter.

First, we quantified the impact of the flux noise on the abundance determination. Using the method described in Sect.~\ref{Sect:Derivation}, we actually reanalysed each observed spectrum a very large number of times with GAUGUIN, by estimating 1000 Monte-Carlo realisations of each spectrum, considering its flux uncertainties. Practically, this means that we built 1000 different versions of every observed spectra by noising their flux, considering their \SNR\ ratio. Then, these 1000 noised realisations were automatically and independently analysed, leading to 1000 Pb abundances for a given star. This procedure then allowed to easily estimate the Pb abundance uncertainty caused by the flux uncertainty.
Hereafter, the published A(Pb) corresponds to
the median  (50${th}$ quantile) of the 1000 Pb abundance distribution (hereafter denoted \QC). The associated upper and lower confidence values are estimated from the 84${th}$ and 16${th}$ quantiles (hereafter \QQ\ and \QS, respectively). Therefore, the Pb abundance dispersion defined by \SMC=(\QQ~- \QS)/2 is our adopted abundance uncertainty caused by the spectra \SNR. This \SMC\ means that there is a probability of 68\% that the lead abundance is found within it. For each star, we note that we provide in the AMBRE catalogue \QQ~and \QS~separately since the error bars are not always symmetric. To our knowledge, this is the first time that such a large number of spectra is chemically analysed so many times to derive one of their abundance and an extremely accurate estimate of its associated error caused by the flux uncertainties. We finally point out that \SMC\ also takes  into account some possible internal errors of the whole procedure as continuum normalisation, spectra quality, radial velocity correction. 

It is found that, among the 1341 spectra selected in Sect.~\ref{SelectSpec}, 4\% and 0.2\% of them have a \SMC\ larger than 0.10 and 0.20~ dex, respectively. We report in Table~\ref{MCTab} the mean estimated uncertainties of \PbM~for three main stellar types (cool giant, cool dwarf and solar-type) and for two values of \SNR~(65 and 110). The presented mean values are computed from all the AMBRE spectra which parameters are within a range of $\pm$~150 K around the adopted central temperature and $\pm$ 15 around the indicated \SNR. 
The number of stars (N$_S$) used for the computation of the presented quantities is also indicated.
As expected, \SMC\ decreases with increasing \SNR, leading to small dispersions for high S/N spectra ($\la$0.05~dex). \\

\begin{table}[t!]
        \caption{\label{MCTab} Uncertainties of the lead abundances  A(Pb) caused by the spectra \SNR~for different stellar types.}
        \centering
        \begin{tabular}{lcccccc}
                \hline
                \hline
                & \multicolumn{2}{c}{Cool giant} &  \multicolumn{2}{c}{Cool dwarf} & \multicolumn{2}{c}{Solar-type} \\
                \T      & \multicolumn{2}{c}{$\sim$ 4500 K} & \multicolumn{2}{c}{$\sim$ 5100 K} & \multicolumn{2}{c}{$\sim$ 5700 K} \\
                \SNR & $\sim$ 65 &  $\sim$ 110 & $\sim$ 65 & $\sim$ 110 &$\sim$ 65 &$\sim$ 110 \\
                \SMC                              & $\pm$ 0.05 & $\pm$ 0.04 & $\pm$ 0.05  & $\pm$ 0.03 & $\pm$ 0.05 & $\pm$ 0.04 \\
                
                N$_{S}$       & 28   & 9    &  17    &  21   & 35   & 50 \\
                \hline
        \end{tabular}
        \tablefoot{N$_S$ is the number of spectra selected to estimate the mean \SMC~for each stellar type.}
\end{table}

Then, we estimated the lead abundance A(Pb) sensitivity to typical uncertainties of the atmospheric parameters determined by the AMBRE Project, by reanalysing every spectrum adopting modified sets of atmospheric parameters. They are presented in Table~\ref{ErrorsTab} separately for a \T, \g, \Meta~uncertainty of $\pm$~120K, 0.25 and 0.10 dex, respectively, which are the mean values reported in the AMBRE parameter tables. We also report their quadratic sum (hereafter denoted \D) for the same three stellar types as in Table~\ref{MCTab}. We emphasise that this \D\ has been computed for each spectrum and is then reported for each star in the AMBRE catalogue.
From Table~\ref{ErrorsTab} we note that \PbM~is more sensitive to temperature uncertainty, especially for cool giants. Even though the lines are more visible in their spectra, the continuum level placement is strongly impacted by the presence of numerous atomic and molecular features
and could lead to less well defined line depths. 

We note that errors of \Vmi~$\pm$ 1~km.s$^{-1}$ and \AF~$\pm$ 0.10~dex have no significant effect on the abundance value. This last point can be explained by the absence of lines from $\alpha$-elements such as TiO in the studied spectral domain. We also point out that the above estimated errors obtained by varying \T~and \g~also include macroturbulent velocity changes as this parameter value is modified according to the relation adopted in this work.  


\begin{table}[t!]
        \caption{\label{ErrorsTab} Uncertainty of the A(Pb)  lead abundances produced by typical uncertainties on the stellar atmospheric parameters for high-quality spectra (\SNR>100).}
        \centering
        \begin{tabular}{lccc}
                \hline
                \hline
                & \multicolumn{1}{c}{Cool giant} &  \multicolumn{1}{c}{Cool dwarf} & \multicolumn{1}{c}{Solar-type} \\
                $\Delta$ \T    = $\pm$ 120 K    & $\pm$ 0.09 & $\pm$ 0.06 & $\pm$ 0.06 \\
                
                $\Delta$ \g    = $\pm$ 0.25~dex     & $\pm$ 0.02 & $\pm$ 0.04 & $\pm$ 0.02 \\
                
                $\Delta$ \Meta = $\pm$ 0.10~dex  & $\pm$ 0.04 & $\pm$ 0.04 & $\pm$ 0.04 \\
                
                \D                  & $\pm$ 0.10 & $\pm$ 0.08 & $\pm$ 0.07 \\
                
                
                
                \hline
        \end{tabular}
\end{table}

\subsection{Final AMBRE:Pb catalogue \label{Catalog}}
As already discussed, our sample is composed of a large number of stars for which several spectra are available. Similarly to the procedure adopted in \citet{Pablo20} and \citet{Perdigon21}, we performed a cross-match between the \Gaia\ e-DR3 ID \citep{GaiaeDR3} and the AMBRE catalogue. For a couple of stars, no \Gaia\ ID were found and we therefore adopted their HD name.

Among the 1341 spectra having passed the selection criteria presented in Sect.\ref{SelectSpec} (e.g. removing highly rotating stars, too low \SNR, no detected lines), we found that they correspond to 653 different stars, 32.1\% of them being giants. 72.4\% of the total sample stars have only one spectrum whereas 3.5 \% of them have more than ten repeats. The metallicity range of this sample star varies from approximately -2.9~dex to ~+0.6~dex with lead abundances varying from -0.7~dex to 3.3~dex.
For each star, we checked that the repeat spectra (if some are identified) were parametrised in a very consistent way  among each other. We indeed systematically rejected individual spectra of a given star whose parameters are found outside the standard deviation in \T, \g~and \Meta, calculated for all the available repeats. The adopted threshold were 110K, 0.5 and 0.20 dex, respectively.\\

We report this AMBRE catalogue of lead abundances in Table~\ref{tab:Catalog}. With about 650 stars, this is the largest catalogue ever published 
(as  can be seen from Fig.~16 and 32 of \citet{Prantzos18, Kobayashi20}, respectively). For each star, we report their number $N_{\rm spec}$ of available repeat spectra if any (only spectra who passed the selection criteria), the star ID (\Gaia\ DR3 or HD name) and their \T, \g, \Meta, \AF\ (or mean values if repeats are available). The lead abundances are provided by the median of the LTE A(Pb) Monte-Carlo distributions~(\QC\ values, or their mean in case of repeats), together with their associated \QQ\ and \QS . The quadratic sum of errors caused by atmospheric parameters uncertainties (\D) as described in Sect \ref{Error}, is also provided.

We then took into account all the uncertainty sources estimated above by computing the quadratic sum of \D\ (caused by the atmospheric parameter uncertainties) and \SMC\ (caused by the spectra \SNR). We found a mean uncertainty ($\Delta$Pb) of about 0.10~dex with a standard deviation of 0.04~dex, for all the AMBRE:Pb catalogue stars.  This uncertainty $\Delta$Pb is published for any star in  Tab.~\ref{tab:Catalog}. We also found that the mean Pb abundance dispersion for stars having more than one repeat is about 0.04 dex. We can therefore conclude that the reported lead abundances present a very good consistency and measurement precision between each other.

Finally, we also report in Table \ref{tab:Catalog} the corresponding NLTE Pb abundances. They were estimated by adding to our LTE lead abundance the NLTE corrections calculated by \citet[][and private communication for the NLTE corrections specific to the selected \PbI\ 368.3nm line]{Mashonkina12}. The applied NLTE corrections were interpolated in the provided table at the parameters of our stars.
They are of the order of $\sim$+0.15~dex for dwarf stars and about twice larger for giants. They satisfactory correct the 
systematic differences observed between the LTE Pb abundances of these two types of stars.\\

To illustrate this AMBRE:Pb catalogue, we show in Fig. \ref{fig:Kiel-Pb} the Kiel diagram of our 653 stars colour-coded by their metallicity (left panel) and their NLTE Pb abundance (right panel). We can see again that the majority of our sample is dominated by dwarf stars. Rather few stars seem to be on the Asymptotic Giant Branch and a few others are metal-poor and lead-enhanced. They are discussed in Sect. \ref{Sect:AGB} and \ref{Sect:LEMP}, respectively.

\begin{figure*}[h!]
        \includegraphics[scale = 0.4]{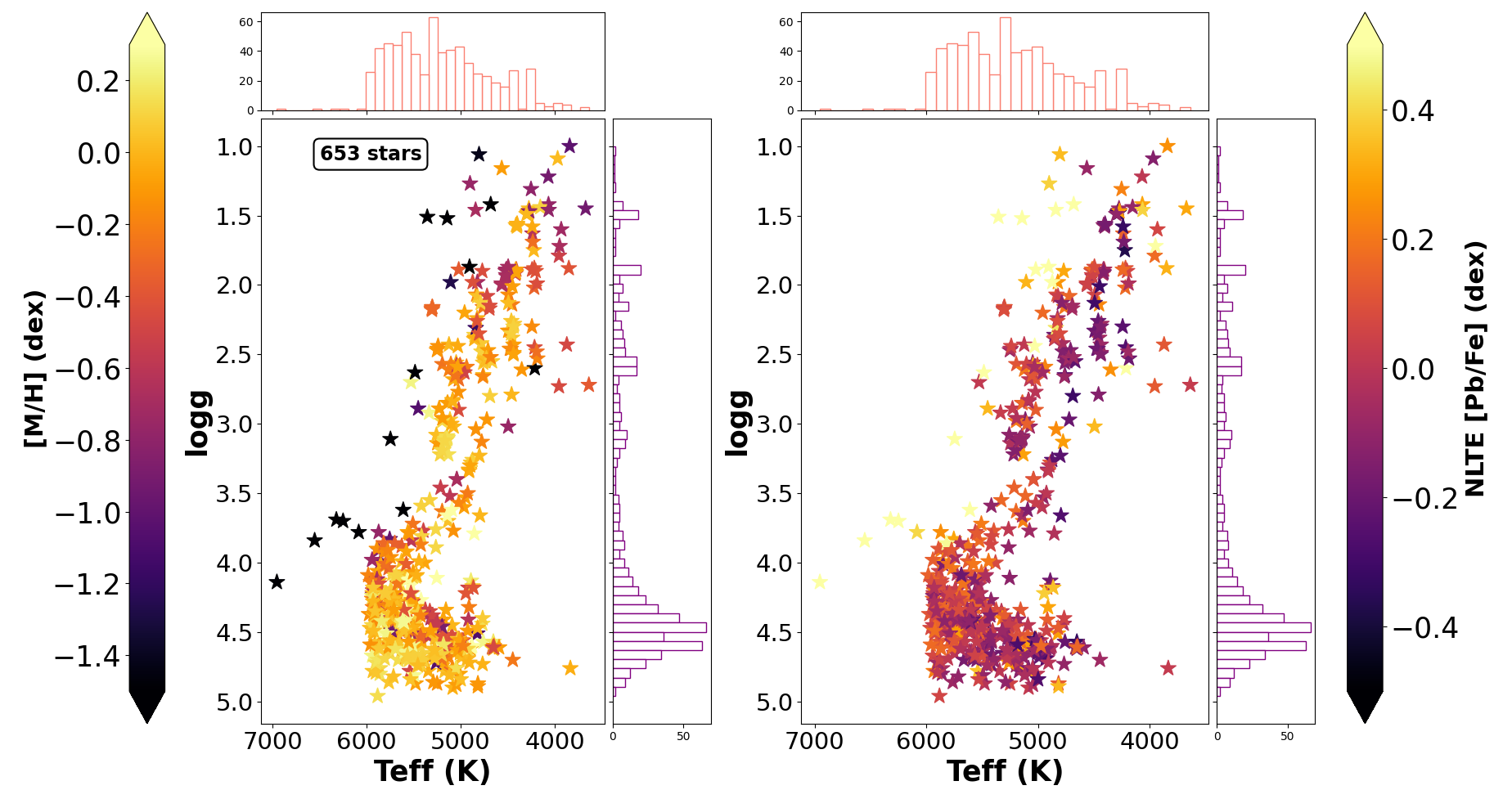}
        \caption{Kiel diagrams of the 653 AMBRE:Pb stars, colour-coded by their metallicity (left panel) and their NLTE Pb abundance (right panel) 
 }
        \label{fig:Kiel-Pb}
\end{figure*}


\begin{table*}[h!]
        \caption{\label{tab:Catalog}  AMBRE catalogue of lead abundances.}
        \centering
        \begin{tabular}{lcccccccccc}
                \hline
                \hline
                Star Name & $N_{\rm spec}$ & \T & \g & \FeH  & \AF & A(Pb) LTE & A(Pb) NLTE &  A$_{Q16}$ & A$_{Q84}$ & $\Delta$Pb \\
             & & K & & dex  & dex & dex & dex & dex & dex & dex \\
             \hline
                HD 124897 & 1 & 4236.3 & 1.63 & -0.63 & 0.15 & 1.23 & 1.52 & 1.19 & 1.26 & 0.14 \\
                HD 29139 & 1  & 3670.9 & 1.45 & -0.93 & 0.00 & 1.06 & 1.38 & 1.00 & 1.11 &  0.06 \\
                11210418694164864 & 1  & 4445.4 & 4.70 & -0.24 & 0.04 & 1.55 & 1.69 & 1.51 & 1.59 &  0.07 \\
                1153464835149837696 & 1  & 5286.2 & 4.51 & -0.78 & 0.17 & 1.15 & 1.33 & 1.09 & 1.20 & 0.08 \\
                1160260989536170880 & 4 & 5722.4 & 4.26 & 0.02 & -0.17 & 1.97 & 2.09 & 1.92 & 2.02 &  0.16 \\
                ... & ... & ... & ... & ... & ... & ... & ... & ... & ... & ... \\
                \hline
        \end{tabular}
        \tablefoot{The full version is available in electronic form. The A$_{Q16}$ and A$_{Q84}$ correspond to the LTE A(Pb) of the 16${th}$ and 84${th}$ quantiles, respectively}
\end{table*}

\begin{figure}[h!]
        \resizebox{9.7cm}{6.5cm}{\includegraphics{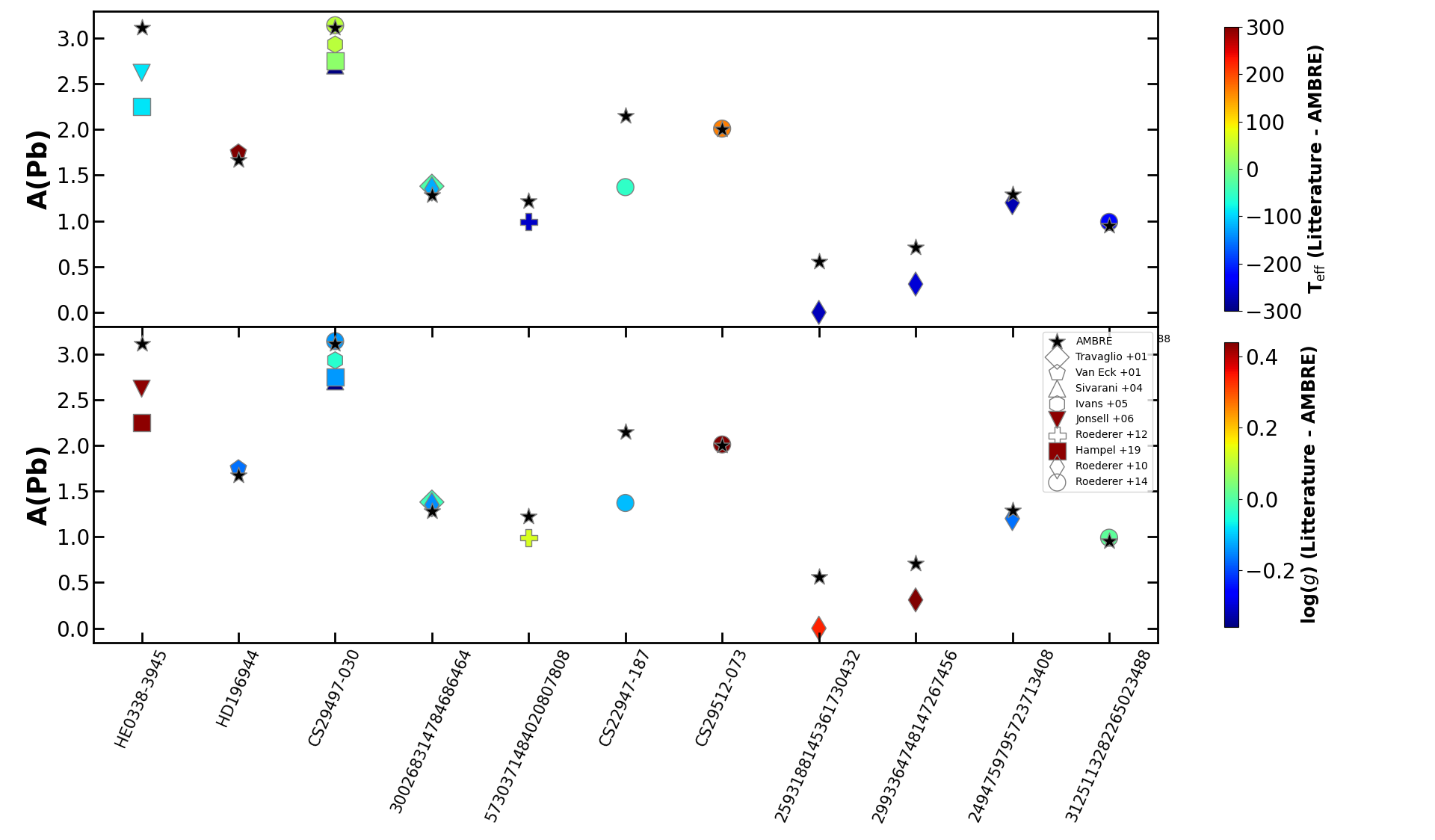}}
        \caption{AMBRE LTE A(Pb) abundances (filled stars) and the literature values (see legend in inset): wide diamonds are from \citet{Travaglio01}, pentagon from \citet{VanEck01}, up pointing triangle from \citet{Sivarani04}, hexagon from \citet{Ivans05}, down pointing triangle from \citet{Jonsell06}, plus sign from \citet{Roederer12}, squares from \citet{Hampel19}, narrow diamonds from \citet{Roederer10}, and circles from \citet{Roederer14}. The top and bottom panels are colour-coded by the difference between the literature and AMBRE \T\ values and the literature and  \g\ values, respectively.}
        \label{fig:PbLit}
\end{figure}

\subsection{Validation thanks to literature comparisons}
Among the 653 stars present in our AMBRE:Pb catalogue, eleven have LTE Pb abundances already published in the literature. Figure~\ref{fig:PbLit} shows the comparison between these AMBRE:Pb abundances and those of the literature colour-coded by the differences in \T~(top panel) and \g~(central panel) and \Meta~(lower panel) in the sense (Literature - AMBRE).

Globally, we first remark that the AMBRE:Pb abundances are compatible with the literature, as are their adopted atmospheric parameters. 
The median of the differences in Pb is around -0.2~dex with a dispersion of 0.26~dex, which corresponds to the typical dispersions between the different literature values, considering their uncertainties. 
Only for two peculiar stars the differences are larger. First, our Pb abundance for HE 0338-3945 (\Gaia\ DR3 4856059422664301440, a very metal poor star with \Meta = -2.09 dex) is larger than that of \citet{Jonsell06} and \citet{Hampel19} by 0.50 dex and 0.86, respectively. These differences in the Pb abundances can be explained by the different values of \g\ and \Meta\ adopted by the different studies. The AMBRE \g~is indeed about 0.4 higher than that of \citet{Jonsell06} and \citet{Hampel19}. But, more importantly, differences in metallicity of about 0.6~dex are found between AMBRE and these two other works. 
Second, for CS 22947-187 (\Gaia\ DR3 6663163201607650688, \Meta~= -2.92 dex), our Pb abundance is 0.80 dex higher than that of \citet{Roederer14}, even though their atmospheric parameters are fully compatible with ours. We have carefully re-examined and re-analysed this AMBRE spectra and the low abundance reported by \citet{Roederer14} is not compatible with our data. \\

Finally, we have also compared our adopted AMBRE stellar atmospheric parameters with those recently published by the \Gaia/DPAC GSP-spec module, in charge
of the analysis of the \Gaia\ Radial Velocity Spectrometer data \citep{GSPspecDR3}. After having applied a very strict filtering based on the quality flags recommended by these authors, 417 stars are found in common between AMBRE and GSP-spec. The median of the \T, \g, \Meta\ differences\footnote{The GSP-spec parameters have been calibrated as a function of \T, as indicated in \cite{GSPspecDR3}.} are found to be rather small and equal to 86~K, -0.03~dex and 0.03~dex, respectively, whereas their Median Absolute Deviation is 186K, 0.18 and 0.10 dex. The AMBRE and GSP-spec have therefore published very consistent stellar atmospheric parameters.

In conclusion, the AMBRE:Pb catalogue provides high-quality stellar parameters and lead abundances. It allows to explore the lead chemical properties in different type of stars (Sect.~\ref{Sect:peculiar}) together with the evolution of this species in the Milky Way (Sect.~\ref{Sect:MW}).

\section{Lead abundance in peculiar stars \label{Sect:Pec}}
\label{Sect:peculiar}

The AMBRE:Pb catalogue contains a total of 653 stars.
 Most of them are main-sequence (429 stars, i.e. 66\%, defined by \g>3.5) and red giant stars (204 stars, 31\%) with metallicities greater than -1.0~dex. 20 stars (of any gravity) are found to be more metal-poor than this value and 15 dwarf are more metal-rich than 0.3~dex.
In this catalogue, one also finds three AGB stars, nine lead-enhanced metal-poor stars and 13 \Gaia\ Benchmark Stars.
Their content in lead is described in the following subsections.

\subsection{Asymptotic Giant Branch stars \label{Sect:AGB}}
In the AMBRE:Pb catalogue, three stars (HD 29139, \Gaia\ DR3 5306951892625773312, and \Gaia\ DR3 6083708032481982336) have atmospheric parameters suggesting that they could be located on the Asymptotic Giant Branch (AGB,  \T~< 4000K and \g~<2.0). These stars are found to be not enriched in Pb, their [Pb/Fe] being close to 0~dex. Therefore, they are probably in the early AGB phase and have not yet experienced third dredge-up events that should have enriched their atmosphere in Pb. This is consistent with their rather hot \T\ (larger than $3700$~K) and rather high surface gravity found between 1.1 and 1.5, with respect to more evolved AGB. We finally recall that, during the AMBRE parametrisation, we assumed that these AGB are not carbon-rich, which is also consistent with their no lead enrichment nature.

\subsection{Lead enhanced metal-poor stars \label{Sect:LEMP}}
In the AMBRE:Pb catalogue, nine metal-poor stars\footnote{None of them being found in the \Gaia/GSP-spec catalogue \citep{GSPspecDR3}} are found to be strongly enhanced in lead, with \PbM$>$1.5~dex, without considering NLTE corrections. Among them, only two have already been identified in the literature as lead-enhanced stars, and therefore, seven are new discoveries:

First, our A(Pb) abundance of \Gaia\ DR3 4856059422664301440 (also known as HE 0338-3945) confirms its Pb-enhanced nature previously reported by \citet{Jonsell06} and  \citet{Hampel19}. However, as already discussed in the previous section, these three reported Pb abundances differ slightly because of differences in the adopted atmospheric parameters. Moreover, this star has been identified as a Carbon-enhanced metal-poor star (CEMP) of s-II type ([Ba/Eu] > 0.5 dex and [Eu/Fe] > 1.0 dex) according to \citet{Barklem05}.  Such enhancements have also been confirmed by \citet{Abate15}. 

Similar conclusions can be drawn for \Gaia\ DR3 6907236881547590144 (HD 196944) whose Pb abundance was also previously derived by \citet{VanEck01}. This star seems also to be a CEMP star classified as CEMP -s/rs according to \citet{Yoon16} or CEMP-s according to \citet{K21}. Such carbon and $s$-process element enhancement have also been highlighted in \citet{Zacs98}.\\

For the seven newly discovered  Pb-enriched stars reported in the present work, we
note that \Gaia\ DR3 6663163201607650688, 2347402354016302208, 2604742840043317760, 2604742840043317760, 3790607572739648000 and  2601354871056014336 were also already identified as CEMP stars with $s$-process enrichments \citep{Aoki02, Komiya07, Abate15, Yoon16}. However, no chemical information have been found for \Gaia\ DR3 6299901567857374848, that could be a new CEMP star candidate. Its parameters are \T=4680~K, \g=1.4, \Meta=-2.6~dex and \PbM$=$1.58~dex.

It is interesting to note that, thanks to \Gaia\ DR3 astrometric information, the binary nature of these stars can be explored. We have indeed check their $ruwe$ and $duplicated\_source$ parameters, their $ipd\_frac\_multi\_peak$ and $non\_single\_star$ parameters showing no peculiar value. Among the nine stars described above, five are suspected binaries that could help to interpret their nature. \Gaia\ DR3  2347402354016302208, 
2604742840043317760, 3790607572739648000 and 6907236881547590144 have their $ruwe$ parameters well larger than 1.4. Moreover,  \Gaia\ DR3 6663163201607650688
has its $duplicated\_source$ equal to $True$. Therefore, following \Gaia, these five stars  of the sample of Pb-enriched metal-poor stars could therefore belong to a binary system whereas the remaining four do not have any clues of a possible multiple star  nature. This should be confirmed with the future \Gaia\ data releases that will span a much larger timescale than the DR3, helping to better detect any possible binary signature.

Several scenarios could be invoked  to produce such Pb enriched stars and, more generally, $s$-process elements enrichments. The main one for CEMP-$s$ stars is a binary-mass transfer between an AGB star that has produced carbon and heavy elements and a less evolved companion 
\citep[e.g.][]{Abate15}. It is highly probable that our candidates belong to this category. However, some stars are both $s$- and $r$-process enriched and could thus not be formed only thanks to AGB winds. Even if not all the processes are understood, many scenarios have been investigated to explain such exotic abundance patterns. Among them, one could first cite a binary system formed in a molecular cloud already enriched in $r$-process elements by the nearby explosion of one or more Type II supernovae \citep[e.g.][]{Jonsell06, Bisterzo12, Sneden14}. Another scenario assumes that one of the member of a binary system is relatively massive and undergoes the AGB phase, producing $s$-elements, but then explodes as an electron-capture supernova, providing
the $r$-elements \citep[e.g.][]{Jonsell06}. 
Finally, many works have suggested that both $r$- and $s$-elements could be synthesised in low-mass, extremely metal poor AGB stars ([Fe/H]$<$-3~dex): these stars might generate high densities of free neutrons during the ingestion of protons in
the region of the He-flash \cite[e.g.][]{Sergio09, Lugaro09, Sergio16, Choplin21, Choplin22}.


\subsection{Lead abundance in \Gaia\ Benchmark Stars \label{GBS}}
Finally, Table \ref{tab:Catalog} also contains lead abundances, adopting the AMBRE atmospheric parameter values, for 13 \Gaia\ benchmark stars, hereafter denoted GBS  \citep[see][and references therein]{Jofre18}.
These FGKM-type stars are commonly adopted to calibrate and/or validate large spectroscopic surveys as their atmospheric parameters were carefully derived using several techniques. 

For some of them, these parameters could slightly differ from the derived AMBRE ones since the analysis technics and observed data differ. For consistency with already published values, we recomputed their lead abundance adopting the atmospheric parameters of \citet{Jofre18}. The results are presented in Table \ref{tab:GBS}, which is structured similarly as Tab.~\ref{tab:Catalog}, (excepted for the \D~column). We note that the adopted \AF~is the mean of the individual Mg, Si, Ca and Ti abundances from \citet{Jofre18}. We also added the solar and Arcturus abundances already shown in Table~\ref{tab:GBS} computed from \citet{Wallace11} and \citet{S4N2004} spectra, respectively. Except for HD120283, for which we found a \PbM~difference of 0.15 dex with respect to Tab.~\ref{tab:Catalog}, caused by a \T~difference of about 220~K, we remark that these new \PbM~abundances reported in Tab.~\ref{tab:GBS} differ by only 0.05 dex (in median) with respect to the AMBRE:Pb values. Such tiny differences are completely explained by the small differences on the adopted atmospheric parameters.

\begin{table*}[h!]
        \caption{\label{tab:GBS} Lead abundances in \Gaia\ Benchmark Stars adopting the \citet{Jofre18} atmospheric parameters.}
        \centering
        \begin{tabular}{llccccccc}
                \hline
                \hline
        Star  & \Gaia\ DR3 or HD  & $N_{spec}$ & \T & \g & \FeH & \AF &  LTE A(Pb) & $\sigma_{MC}$  \\
   &   &  & K &  & dex & dex &  dex & dex  \\
        \hline
        HD140283    & 6268770373590148224 &  1  & 5522 & 3.58 & -2.36 &   0.03 &  1.06 &  0.13\\
18~Sco      & 4345775217221821312 &  4  & 5810 & 4.44 &  0.03 &   0.02 &  2.00 & 0.03   \\
$\alpha$ Cen A  &                       HD128620                  &  2  & 5792 & 4.31 &  0.26 &  -0.04 & 1.79 & 0.07  \\
$\alpha$~Cen B  & HD128621                        &  3  & 5231 & 4.53 &  0.22 &   0.08 & 1.80 & 0.07   \\        
$\alpha$ Tau  &                 HD29139                   &  1  & 3927 & 1.11 & -0.37 &   0.11 & 1.75 & 0.13  \\
$\beta$ Hyi   &  4683897617110115200  &  3  & 5873 & 3.98 & -0.04 &  -0.02 & 1.97 & 0.04     \\    
$\epsilon$~Eri  & 5164707970261890560 & 210 & 5076 & 4.61 & -0.09 &   0.03 &  2.20 &  0.03\\       
$\epsilon$~For  & 5071514326764428544 &  4  & 5123 & 3.52 & -0.60 &   0.33 & 1.90 & 0.04 \\ 
$\epsilon$~Vir  & 3736865265441207424 &  1  & 4983 & 2.77 &  0.15 &  -0.07 &  2.10 &  0.10 \\
        $\mu$~Ara   & 5945941905576552064 &  1  & 5902 & 4.30 &  0.35 &   0.00 & 1.99 & 0.05 \\
$\tau$~Cet      & 2452378776434477184 & 13  & 5414 & 4.49 & -0.49 &   0.23 & 1.97 & 0.02 \\ 
          Sun\tablefootmark{a}  &       -         &  1  & 5771 & 4.44 & 0.00  &   0.00 & 1.80 & 0.01 \\
Arcturus\tablefootmark{b}  &                    HD124897                  &  1  & 4286 & 1.60 & -0.52 &   0.24 & 1.86 & 0.05 \\
Arcturus\tablefootmark{c}  &                    HD124897                  &  1  & 4286 & 1.60 & -0.52 &   0.24 & 1.62 & 0.05    \\
                \hline
        \end{tabular} 
\tablefoot{
        \tablefoottext{a}{\citet{Wallace11} spectrum,}
        \tablefoottext{b}{AMBRE UVES spectrum}
        \tablefoottext{c}{S4N (\citet{S4N2004}) spectrum}.}
\end{table*}

\section{The Milky Way lead abundance} 
\label{Sect:MW}

\subsection{Spatial and kinematic distributions}

In this section, we first investigate the spatial and kinematical distribution of the AMBRE:Pb Catalogue.
For that purpose, we computed the stellar positions (Galactocentric Cartesian coordinates X, Y, Z) as well as the Galactocentric radius ($R$) and cylindrical velocities (V$_R$ , V$_Z$ and V$_\phi$) as presented in \citet{PVP_Ale}, using the \Gaia\ eDR3 photo-geometric distances from \citet{Coryn21}.  We note that these quantities were computed for the 485 AMBRE:Pb stars having \Gaia\ DR3 data, especially distances (74\% of the whole sample).
The top panels of Fig.~\ref{fig:XYRZ} show this sample in the (X, Y) plane, whereas the bottom panels correspond to the ($R$, Z) plane. The left panels are colour-coded by the stellar counts and the right panels by V$_\phi$. It can be seen that the majority of the stars (92.7 \%) are located in a box of 1~kpc size around the solar position, confirming the solar neighbourhood nature of the AMBRE:Pb sample. This is confirmed by the right panels of Fig. \ref{fig:XYRZ} where 91\% of the stars are found close to the Galactic Plane (|Z| < 0.7~kpc).

Moreover, Fig. \ref{fig:Toomre} show the Toomre diagram of these 485 stars. 70\% of the stars have a total velocity smaller than 70~km.s$^{-1}$, an indication that they may belong to the disc component of the Milky Way. However, some metal-poor stars present a total velocity larger than 150~km.s$^{-1}$ and may probably belong to the Galactic halo.

\begin{figure*}[t]
        \centering
        \includegraphics[scale = 0.32]{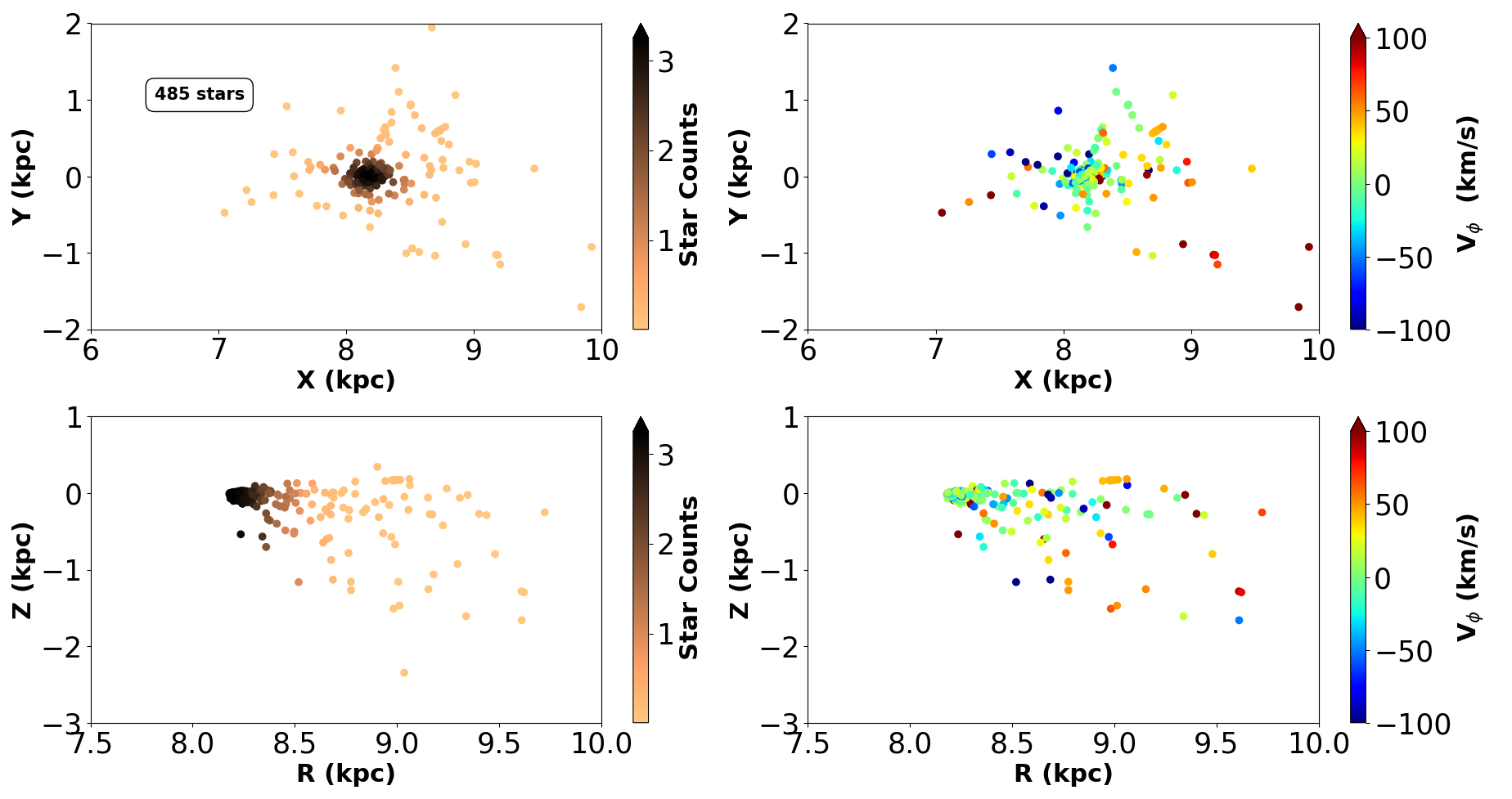}
        \caption{Galactic distributions of the AMBRE:Pb catalogue stars. The top panels show the distributions in Cartesian coordinates (X, Y), colour-coded by stellar counts (left panel) and V$_\phi$ (right panel). The bottom panels show the (R, Z) distributions with the same colour-coding as the top panels. The very few most distant stars are not seen in these plots. }

        \label{fig:XYRZ}
\end{figure*}

\begin{figure}[t]
        \centering
        \includegraphics[scale = 0.22]{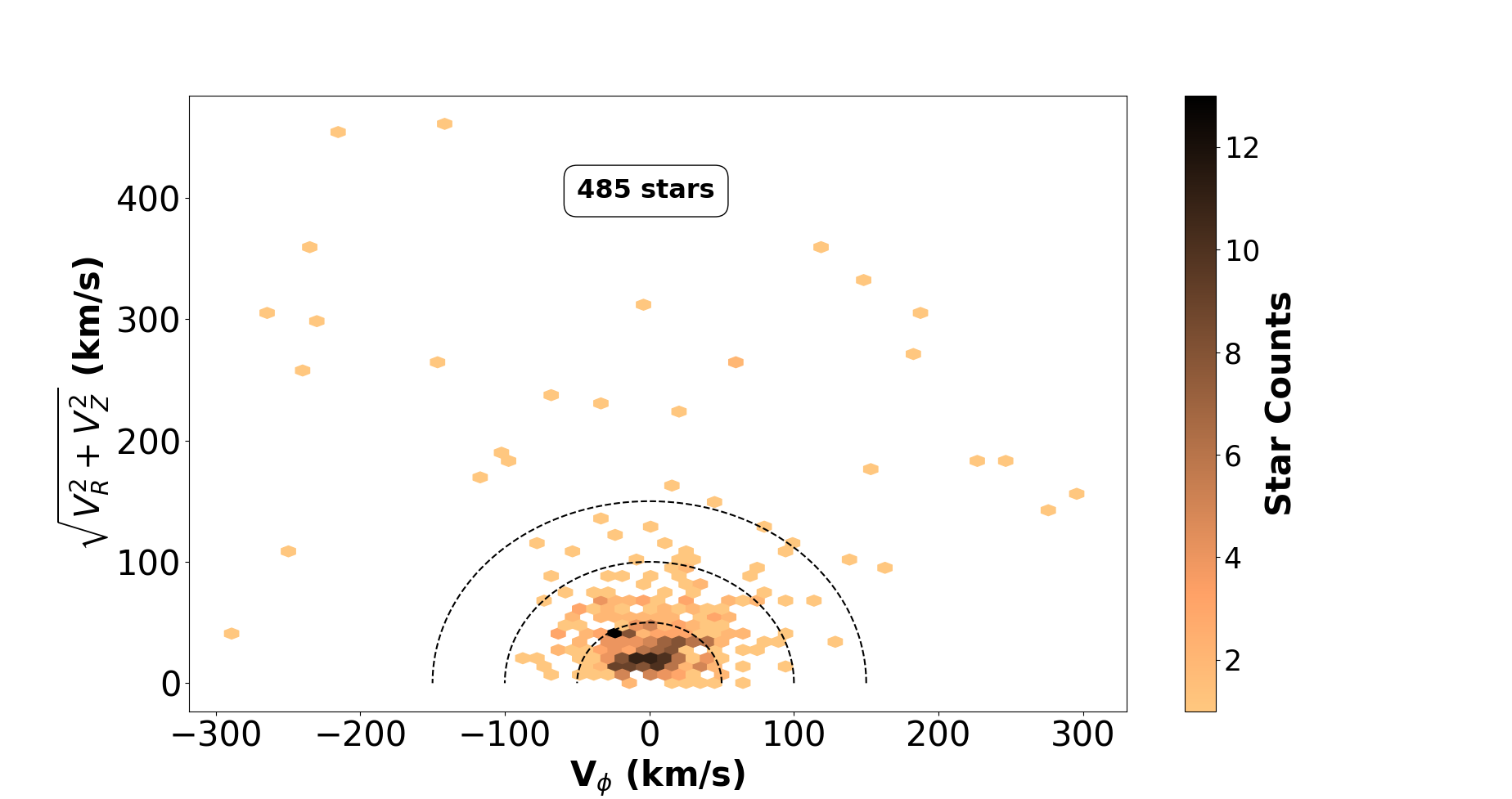}
        \caption{Toomre diagram of the AMBRE:Pb stars colour-coded by their number. 
 }
        \label{fig:Toomre}
\end{figure}


\subsection{Lead in the Milky Way disc}
We then investigate the chemical trend of lead abundances in the Milky Way disc. To do this, we selected only stars with a total velocity smaller than 75~km.s$^{-1}$, |Z| < 0.70~kpc and an uncertainty on Pb abundances smaller than 0.10 dex. We also rejected the above discussed nine Pb-enriched metal-poor stars since their atmosphere is not representative of the ISM chemical composition from which they formed. This selection leads to a sample of 265 $bona-fide$ high-quality disc stars. 
From the selected sample, we computed the Pb radial and vertical gradients of the Galactic disc,
thanks to Theil-Sen fits to the data as already performed in \cite{GOATURSICe}. Uncertainties were computed by adopting a confidence level of 0.95.
We note that, contrarily to previous AMBRE studies \citep[see, for instance,][]{Perdigon21, Pablo21}, we did not separate the thin and thick components of the disc because of too low statistics.

Regarding the [Pb/Fe] radial gradient over a range of $R$ between 8 to 9.5~kpc, we found $\delta$[Pb/Fe]/$\delta$R = 0.012$\pm 0.046$~dex.kpc$^{-1}$. 
This value if fully compatible within error bars with the cerium\footnote{Ce being a second-peak $s$-process element.} radial gradient computed both in \citet{GOATURSICe} and with the value provided by \citet{Taut21}. 
We also found $\delta$[Pb/H]/$\delta$R  = 0.005$\pm 0.017$ dex.kpc$^{-1}$, which is again compatible with the $\delta$[Ce/H]/$\delta$R  derived in \citet{GOATURSICe}, as well as [Ce/H] and [La/H] radial gradients from Cepheids derived in \citet{DASilva16} over a wider range of Galactocentric distances (4–18 kpc).

Then, we estimated the vertical gradients for the Pb abundances for Z-values varying from -0.65 to +0.21~kpc. A positive (but not significative within error bars) trend was found with $\delta$[Pb/Fe]/$\delta$Z = 0.036$\pm 0.037$~dex.kpc$^{-1}$,} similarly to the Ce abundances trend derived by \citet{GOATURSICe}. Similarly, we also derived $\delta$[Pb/H]/$\delta$Z  = 0.010$\pm 0.014$~dex.kpc$^{-1}$.                                    

\subsection{Comparison between the lead, europium, barium and $\alpha$-trends}
\label{sec: Comparison of the lead, europium and alpha-trends}

We now investigate the respective behaviours of the Pb, Eu, Ba and $\alpha$ abundance trends as a function of metallicity in the Milky Way disc (see Fig.~\ref{fig:PbEu}). 
The [Pb/$\alpha$], [Pb/Ba] and [Pb/Eu] ratios were estimated by adopting the lead abundances derived in the present study, the $\alpha$-abundances derived within the AMBRE Project during the stellar parametrisation process \citep{FEROS, UVES}, and the AMBRE europium and barium abundances from \citet{Guiglion18}. For that purpose, we searched which of our 265 Galactic disc stars with high-quality Pb abundances are also found in the Eu and Ba samples, considering only stars with an error in [Eu/Fe] and [Ba/Fe] smaller than 0.10 dex. We end up with 93 and 127 stars with both Pb plus Eu and Ba, respectively. Their metallicities is found in the range between $\sim$-0.8 -- $\sim$0.3~dex.

In Fig.~\ref{fig:PbEu}, we show the different ratios derived for each star (blue points) together with a running mean, adopting bins of 0.1~dex in metallicity (orange points).\footnote{Except for the most metal-rich bin, which corresponds to 0.25~dex for statistics reason.} It can first be seen that [Pb/M] well decreases with the metallicity, reaching [Pb/M] values close to 0 at solar metallicity. The decrease seems to be also present for super-solar metallicities.  Such a decrease is significant, considering the dispersions associated to the running mean (vertical lines for each orange dot).
On the other hand, the [Pb/Eu] ratio seems to be slightly decreasing as metallicity increases. However, the large dispersions associated to these data   are fully compatible with a flat trend around the zero-value. On the contrary, flat trends for [Pb/$\alpha$] and [Pb/Ba], close to 0.0~dex, are shown in the bottom panels, although the dispersions are again rather large. 
These trends are interpreted below thanks to Galactic chemical evolution models.
\begin{figure*}[t]
        \centering
        \includegraphics[scale=0.40]{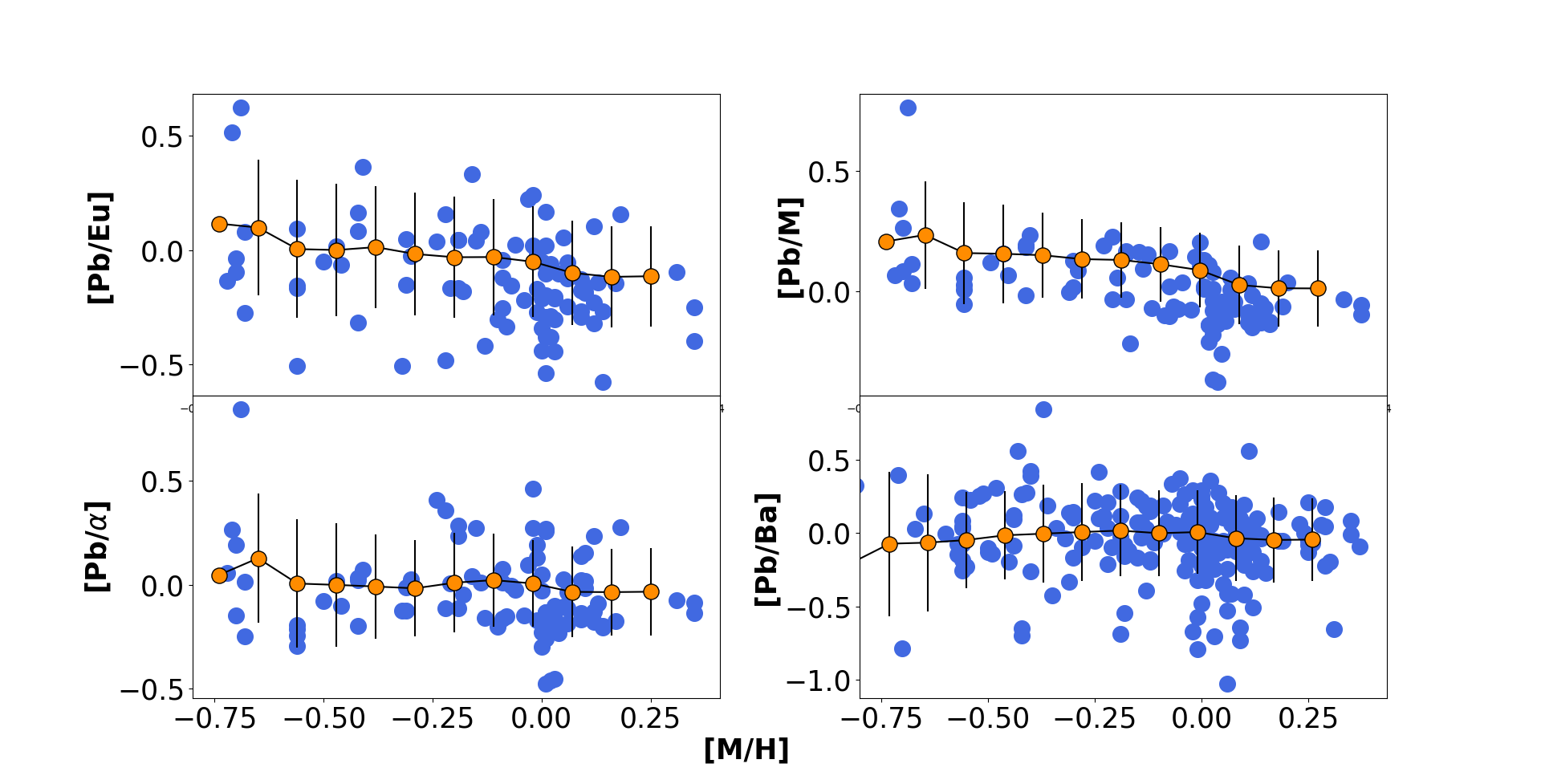} 
        \caption{Behaviour of Pb over Eu, mean metallicity, and $\alpha$ and Ba abundances vs metallicity. 
 The blue points refer to the analysed stars, whereas the orange points are the abundance ratio mean by 0.1~dex metallicity bins (except for the most metal-rich bin which is 0.25~dex). The associated dispersions in each bin is also indicated (vertical black lines).}
        \label{fig:PbEu}
\end{figure*}

\subsection{The Galactic chemical evolution of lead}
The chemical properties of the 265 Galactic disc stars with high-quality Pb abundances are compared to Galactic chemical evolution models, in order to better understand the chemical evolution of lead. This is illustrated in Fig.~\ref{fig: models chemo} where the observed Pb abundance patterns in the [Pb/Fe] versus [M/H]
plane are compared to different chemical evolution models. The chemical evolution models differ between themselves mainly by the number of infall episodes. In particular, the shown models are as follows:

\begin{itemize}
    \item the one-infall model (see e.g. \citealp{MatteucciFrancois1989, Grisoni2018}), which assumes that the solar neighbourhood is formed by means of a single infall of primordial gas with a timescale of $\mathrm{\tau \simeq 7\ Gyr}$.
    \item the two-infall model of \citet{Spitoni2021} (see also \citealp{Spitoni2019}) assuming that the Milky Way disc formed as a result of two distinct gas accretion episodes. The first one formed the chemically defined thick disc (namely the high-$\alpha$ sequence) 
    whereas the second one, delayed by $\mathrm{\sim 4\ Gyr}$, is responsible for the formation of the  thin disc one (namely the low-$\alpha$ sequence). The first gas infall is characterised by a much shorter accretion timescale ($\mathrm{\tau_1\simeq0.1\ Gyr}$) than the second one ($\mathrm{\tau_2\simeq4.0\ Gyr}$).  The original model adopted the set of yields of \citet{francois2004} for massive stars. Here, we  adopt yields derived from \citet{Limongi2018}'s sets, as described below.
    \item the three-infall model of \citet{Spitoni2023} which is an extension of the two-infall one of \citet{Spitoni2021}. It was designed to reproduce the low-$\alpha$ sequence by means of two distinct gas infall episodes (the most recent one which started $\mathrm{\sim2.7\ Gyr}$ ago) in order to reproduce the young chemical impoverishment in metallicity with low [$\alpha$/Fe] values first revealed by the \citet{PVP_Ale}, as well as the recent enhanced star formation activity of \citet{Ruiz-Lara2020}.
\end{itemize}

Prescriptions for the adopted star formation (here parametrised as proportional to the Schmidt-Kennicutt law (\citealp{Kennicutt1998}) and initial mass function (\citealp{Kroupa1993}) are the same for the different models. Moreover, all of them are able to well reproduce the Type Ia SNe, Type II SNe and merging neutron star (MNS) rates as observed by \citet{Capellaro1999} (for SNe) and \citet{Kalogera2004} (for MNS). 

\begin{table}
\caption{\label{tab:chem models}Nucleosynthesis prescriptions for the s-process and number of infalls for different models.}
\centering
\begin{tabular}{lccc}
\hline
\hline
Model  & $V_{rot}$~(km.s$^{-1}$)  & AGB stars & number of infalls \\
\hline
Model 1 & 150 & FRUITY original & one \\ 
Model 2 & - & FRUITY test & - \\ 
\hline
Model 3 & var - DIS1 & FRUITY test & one \\ 
Model 4 & var - DIS2 & - & - \\ 
\hline
Model 5 & var - DIS2 & FRUITY test & two \\
\hline
Model 6 & var - DIS2 & FRUITY test & three \\
\hline
\end{tabular}
\end{table}

With regard to the nucleosynthesis prescriptions, we recall that Pb is formed both through $s-$ and $r-$process channels ($\sim83\%$ and $\sim17\%$ at solar metallicity, respectively according to \citealp{Prantzos2020}). In this work, we adopt a similar nucleosynthesis to that used by \citet{Molero2023}, and we refer to this work for a detailed discussion. In summary, we assume that the $r$-process nucleosynthesis takes place in (i) MNS, with yields scaled to those of Sr from the kilonova AT2017gfo (\citealp{Watson2019}) and in (ii) magneto-rotational supernovae (MR-SNe), with yields from \citet{Nishimura2017}'s model L0.75. MNS are assumed to merge with a delay time distribution (DTD) function and MR-SNe are assumed to be 20\% of massive stars with initial mass between $\mathrm{10-25\ M_\odot}$. The $s$-process nucleosynthesis is assumed to happen in (i) low- and intermediate-mass stars (LIMS; $\mathrm{M\leq8\ M_\odot}$) during their AGB phase with yields prescriptions from the FRUITY database (\citealp{Cristallo2009, Cristallo2011, Cristallo2015}), and in (ii) rotating massive stars ($\mathrm{M\geq13\ M_\odot}$), with yields from \citet{Limongi2018}'s recommended set R. Details regarding the model prescriptions are reported in Table \ref{tab:chem models}. As discussed in Sect.~\ref{sec:Introduction}, the main stellar nucleosynthesis sites of Pb are, in particular, low-mass low-metallicity AGB stars. The non-rotational set of the FRUITY database provides accurate yield grids for 96 stellar progenitors with initial masses and metallicities spanning the ranges $1.3-6.0\ \mathrm{M_\odot}$ and $2\times10^{-5}-2\times10^{-2}$, respectively. Although the great precision provided by a large number of progenitors, the current set of FRUITY yields tends to overestimate the $s$-process contribution for most neutron-capture elements. This feature has already emerged in previous chemical evolution studies (e.g. \citealp{Rizzuti2019, Magrini2021, Rizzuti2021, Molero2024}). The need to reduce the AGB yields appears evident in the case of Pb as well, independently from prescriptions regarding massive stars. This is shown in the upper left panel of Fig. \ref{fig: models chemo}, where Model 1, with no reduction in the AGB yields, creates a bump at intermediate metallicities ($\mathrm{-0.9\leq[M/H]\leq-0.1\ dex}$) dex which overproduces the observed trend. \\As a "Gedanken Experiment" we run a GCE model in which the evolution of the AGB models has been frozen as soon as they attain C/O=1 on their surfaces. As a consequence, $s$-process yields result decreased, being this effect more evident at lower metallicities, where the C-rich regime is reached in the first part of the AGB track. Fig. \ref{fig: models chemo} demonstrates that our working hypothesis, despite being extreme, works in the right direction (Model 2), possibly highlighting that the adopted mass-loss rate in FRUITY models is underestimated. Currently, AGB mass-loss rate in FRUITY models is evaluated basing on observational period -- mass-loss data (see \citealp{Straniero2006} for details). However, the period is theoretically estimated basing on period-luminosity relations in which a distinction between O-rich and C-rich stars has not been operated (e.g. \citealp{Feast1989}). More recent works pointed out that those relations may vary with the C/O ratio, leading to larger mass-loss rates in the C-rich regime (\citealp{Groenewegen2020, Ita2021}). We postpone to a future work a detailed analysis on this topic. Here, we simply conclude that a reduced production of Pb from AGB stars noticeably improves the agreement with the observations. \\
For rotating massive stars, differently from \citet{Molero2023}, two more sets have been considered, corresponding to two different theoretical distributions for the velocities. Similarly to \citet{Romano2019} (see also \citealp{Rizzuti2021}), we assume that the probability that a star rotates faster increases with decreasing metallicity. In this way, we should be able to provide a better fit to the few data observed at $\mathrm{[M/H]\leq-0.1\ dex}$. We recall that the assumption of having a distribution of rotational velocity rather than a fixed one is more physical and supported both theoretically (\citealp{Frischknecht2016}) and observationally (\citealp{Martayan2007, Hunter2008, Chiappini2011, Prantzos18}). The two new distributions, DIS1 and DIS2, assume that stars have an initial rotational velocity of $\mathrm{300}$ and of 175~km.s$^{-1}$, respectively, for $\mathrm{Z\leq3.236\times10^{-3}}$ and of 150~km.s$^{-1}$ for higher values of $\mathrm{Z}$ (Model 3 and Model 4, respectively). As expected, rotation in massive stars increases the production of Pb, causing Model 3 to overestimate the [Pb/Fe] versus [M/H] trend. Only Model 4, with a mild increase in the rotational velocities at low Z, is able to provide an adequate fit with the observations.

Deviations between the Model 4 predictions and the observed trend are present for metallicities above solar for the one-infall model. An agreement in this range of metallicities is partially obtained with a scenario of disc formation including more than one infall (Model 5 and Model 6), in particular in the case of the two-infall model from \citet{Spitoni2021}, because of the shorter first infall timescales (see lower panels of Fig. \ref{fig: models chemo}).
To strengthen our assumptions on the nucleosynthesis, in particular of AGB and massive stars, here we show that our model correctly reproduces the abundance patterns of the second peak $s$-process element Ba. In Fig. \ref{fig: PbBa_BaFe one and two}, we present results for the one-infall and two-infall (from \citealp{Spitoni2021}) models compared to the observed average trends of the [Pb/Ba] and [Ba/Fe] versus [M/H]. We recall that the models shown in the figure differ only by the initial rotational velocities of massive stars (see Table \ref{tab:chem models}). The models are in excellent agreement with the observed trend, in particular for [M/H]$\leq-0.5\ \mathrm{dex}$ and in the case of the [Ba/Fe] versus [M/H]. The [Pb/Ba] ratio is slightly overproduced in the low-metallicity regime by all models, even if the predictions fall well in the uncertainty ranges. 

In Fig.~\ref{fig: models chemo ab average}, we report results for the one-infall and two-infall (from \citealp{Spitoni2021}) models compared to the observed average trends of the [Pb/Eu], [Pb/$\alpha$], [Eu/$\alpha$], and [Eu/Fe] versus [M/H] presented in the previous section. We decided to use oxygen as a comparison for the $\alpha$, since it is actually mainly produced by Type II SNe with low pollution from Type Ia SNe, which makes the comparison easier. All the models are able to provide a good fit with the observed data. The flat trend discussed in Section \ref{sec: Comparison of the lead, europium and alpha-trends} is well reproduced and differences between the models for the [Pb/Eu] and the [Pb/$\alpha$] trends are visible only at low [M/H]. The slight decrease observed in the [Pb/Eu-$\alpha$] as well as the rise of the [Eu/$\alpha$] towards higher [M/H] is also visible in the model predictions, which provide an overall nice agreement with the observations. At low metallicities, the model which best reproduces the observed patterns is the one with higher rotational velocity at low Z (Model 3, corresponding to 300~km.s$^{-1}$) which, as already discussed, increases the production of Pb. It should be noted that higher rotational velocity of massive stars does not affect the production of Eu and of $\alpha$. In fact, in our models Eu is not produced by normal core-collapse SNe. On the other hand, the production of $\alpha$ is increased for rotational velocities higher than 0~km.s$^{-1}$, but above 150~km.s$^{-1}$ there are no significant differences (see \citealp{Romano2019} for details).

\begin{figure*}
\begin{center}
 \subfloat{\includegraphics[width=1\textwidth]{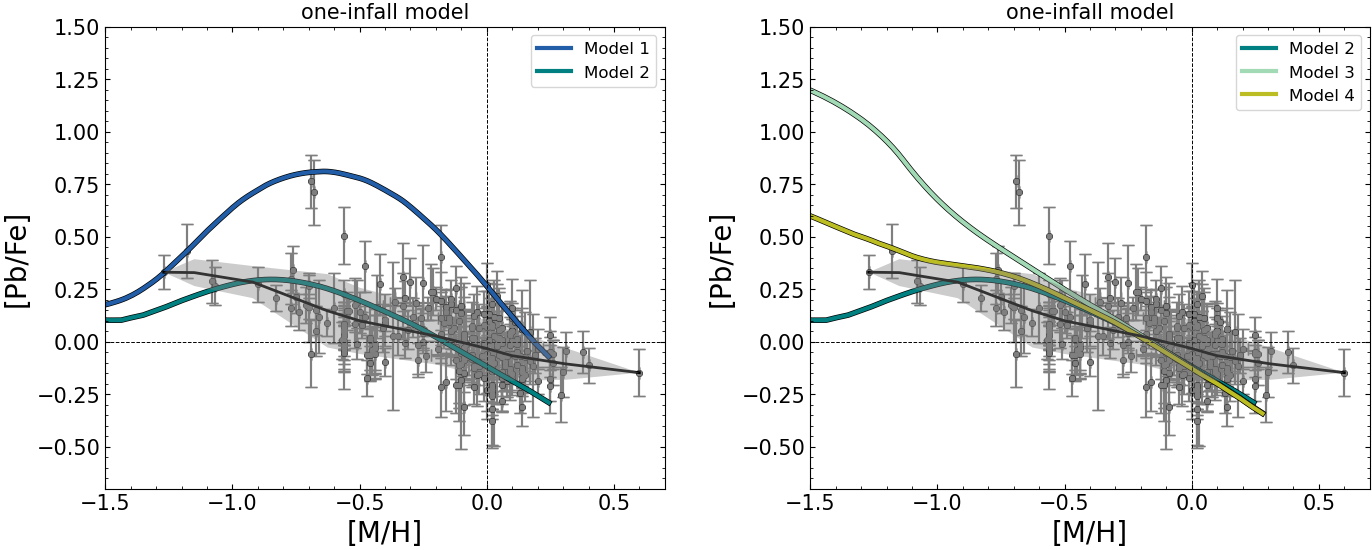}}
 \vfill
 \subfloat{\includegraphics[width=1\textwidth]{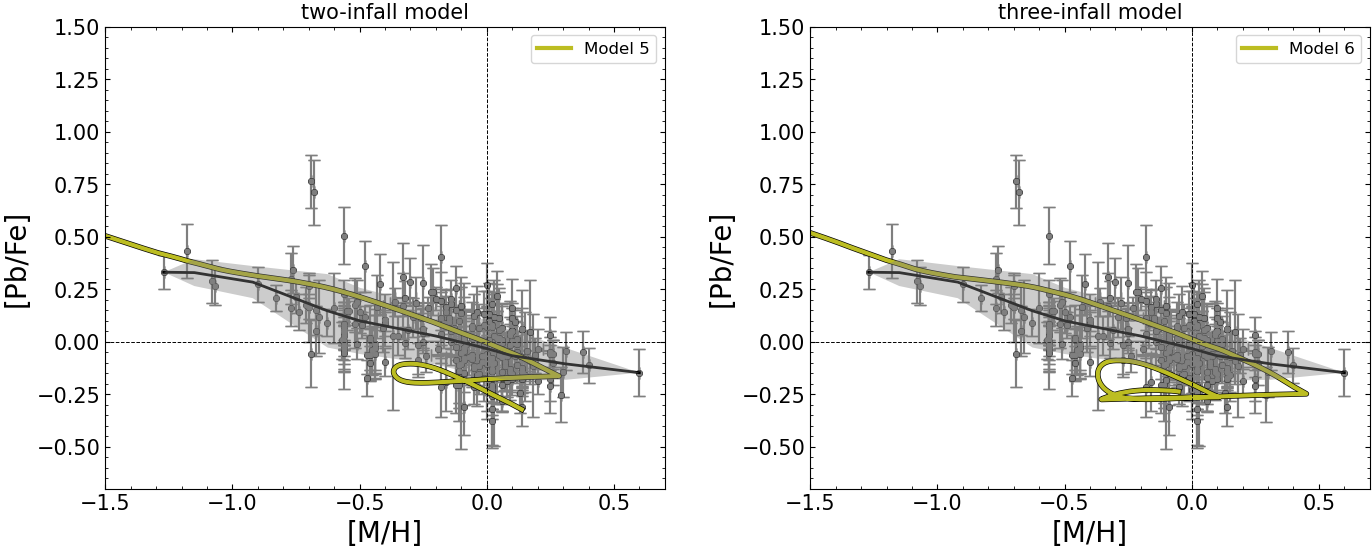}}
 \caption{Model predictions for the [Pb/Fe] vs [Fe/H] abundance pattern. Upper panels: Predictions for a single infall and for different contributions from AGB stars (Model 1 and Model 2) as well as different initial rotational velocities (IRV) of massive stars (Model 3 and Model 4). Lower panels: Predictions for models with more than one infall, massive stars yields weighted with the DIS2 distribution of rotational velocities, and new yields from AGB stars (Model 5 and Model 6). For the models details we refer to Table \ref{tab:chem models}.}%
 \label{fig: models chemo}%
\end{center}
\end{figure*}

\begin{figure}
\begin{center}
\subfloat{\includegraphics[width=1\columnwidth]{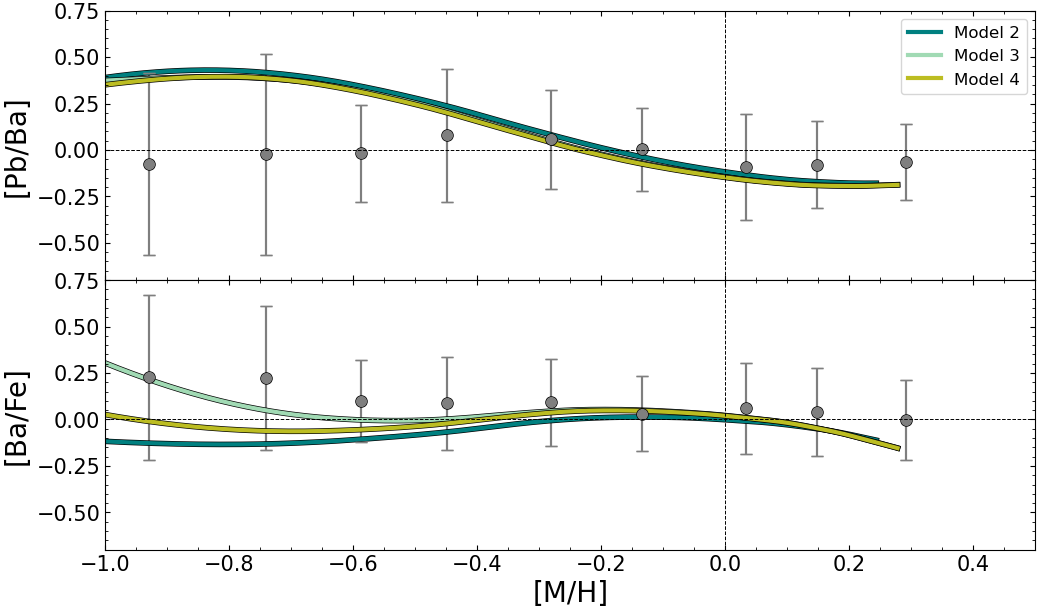}}
\vfill
\subfloat{\includegraphics[width=1\columnwidth]{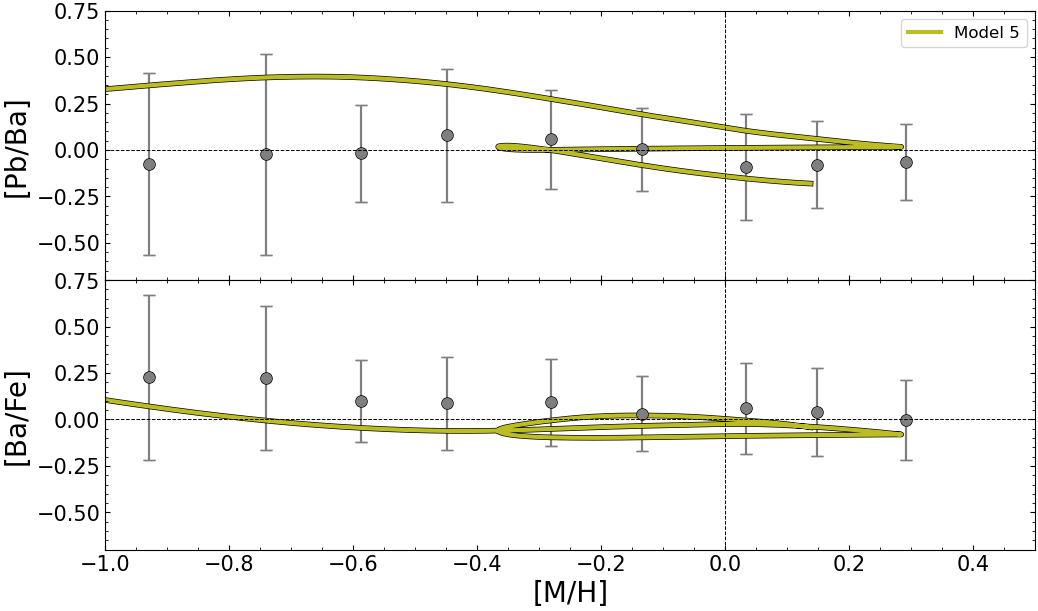}}
 \caption{Model predictions for the [Pb/Ba] and [Ba/Fe] abundance patterns for the one-infall model (upper panels) and the \citet{Spitoni2021} two-infall model (lower panels).}%
 \label{fig: PbBa_BaFe one and two}%
\end{center}
\end{figure}

\begin{figure}
\begin{center}
\subfloat{\includegraphics[width=1\columnwidth]{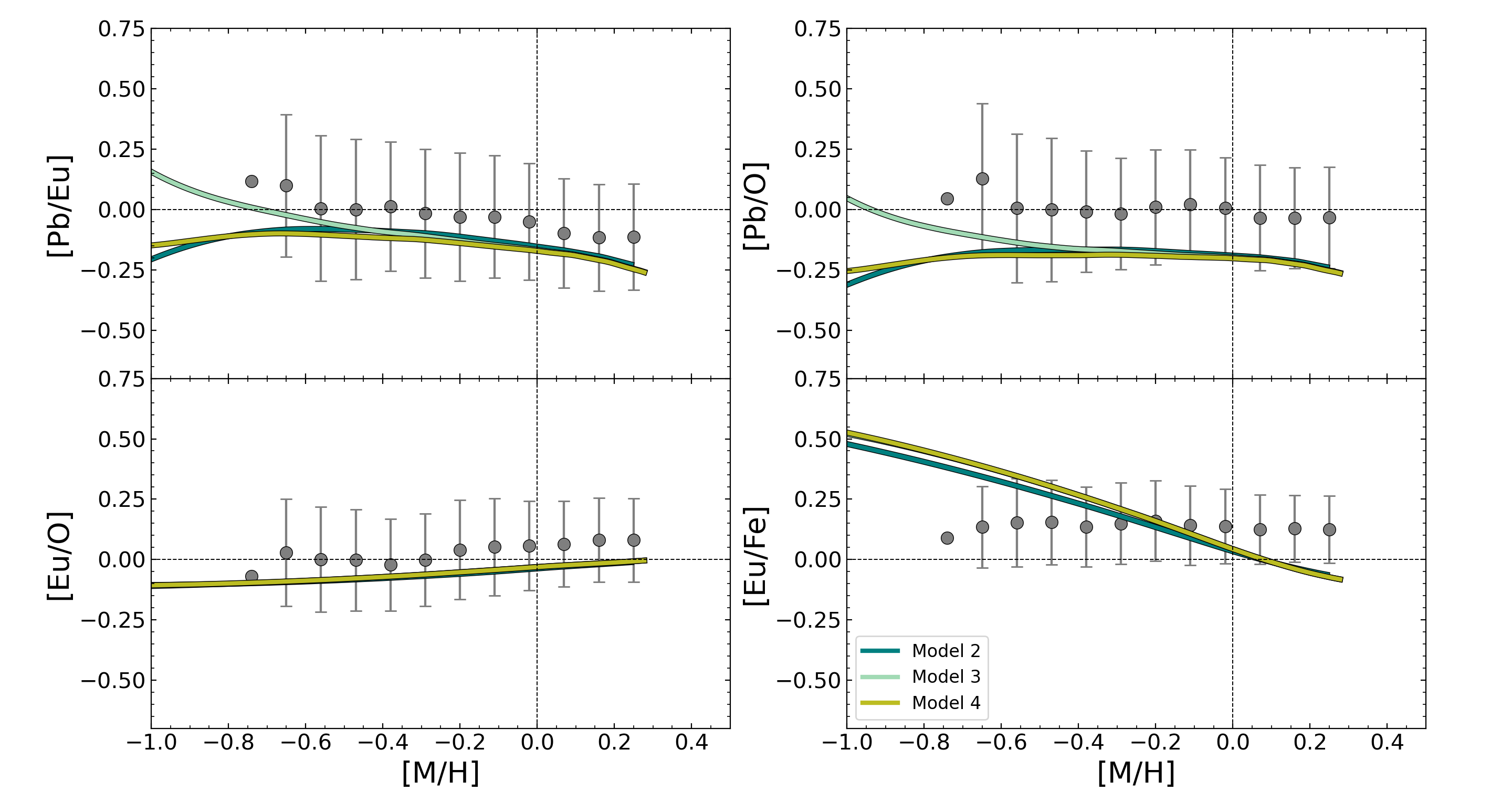}}
\vfill
\subfloat{\includegraphics[width=1\columnwidth]{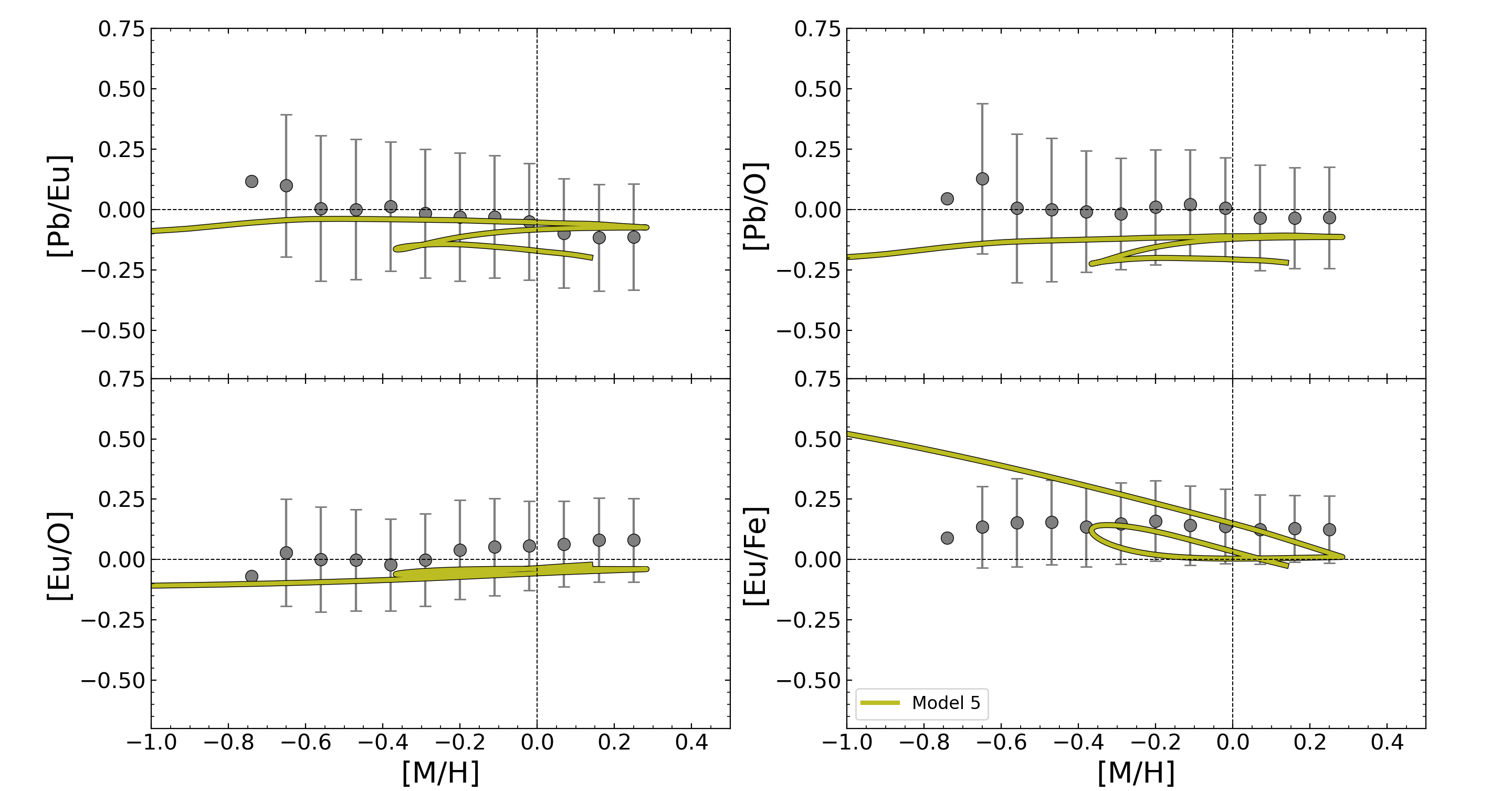}}
 \caption{Same as Figure \ref{fig: PbBa_BaFe one and two}, but for [Pb/Eu], [Pb/O], [Eu/O], and [Eu/Fe] vs [M/H].}%
 \label{fig: models chemo ab average}%
\end{center}
\end{figure}

In summary, the model which better fits the observed [Pb/Fe] versus [M/H] abundance pattern is the two-infall model from \citet{Spitoni2021}, with a distribution of initial rotational velocities of massive stars which favours higher velocities towards lower metallicities. Yields of Pb from AGB stars should be lower than the original FRUITY models, at least for lower metallicities, similarly to what is artificially done for other $s$-process elements (see e.g. \citealp{Rizzuti2019, Molero2023}). The observed trends of the [Pb/Eu], [Pb/$\alpha$], [Eu/$\alpha$], and [Eu/Fe] versus [M/H] are also well reproduced by models that favour higher rotational velocities towards lower [M/H] values. In this case, adopting a one-infall or a two-infall model does not produce significant differences in the predictions.

\section{Summary \label{Sect:Ccl}}
We have presented the AMBRE:Pb catalogue that contains LTE and NLTE lead abundance of 653 stars. This is the largest catalogue of homogeneous lead abundances ever published. It relies on the automatic analysis of the Pb~I~368.346~nm line of more than 8 million spectra, thanks to the GAUGUIN method. The stellar atmospheric parameters were adopted from the AMBRE parametrisation. Pb abundance uncertainties resulting from the \SNR\ spectra were estimated by independently analysing 1,000 flux spectra realisations. This catalogue has then been validated thanks to literature comparison.

Most of the AMBRE:Pb stars are dwarf with metallicities higher than $\sim$-1.0~dex, up to $\sim$+0.3~dex.
Their spatial and kinematic properties confirm that most of them are located within $\sim$1~kpc from the Sun, and probably belong to the Galactic disc. The [Pb/Fe] abundance ratio is found to slightly decrease with metallicity. Radial and vertical gradients of 
$\delta$[Pb/Fe]/$\delta$R = 0.012$\pm0.046$ dex.kpc$^{-1}$
and 
$\delta$[Pb/Fe]/$\delta$Z = 0.036$\pm 0.037$ dex.kpc$^{-1}$
were estimated for disc stars. Moreover, for a sub-sample of these stars, $\alpha$
and europium abundances were previously derived within the AMBRE Project. Flat trends within uncertainties in [Pb/Eu] and [Pb/$\alpha$] over the [-0.75, +0.25] metallicity regime are seen. 

These behaviours have been interpreted thanks to Galactic evolution models. 
The observed trends are well reproduced by the two-infall model, considering either Pb, Eu and $\alpha$-abundance ratios. However, a strong reduced contribution from AGB stars and higher rotational velocities for low-metallicity massive stars have to be invoked. In any case, below \Meta$\sim$-1.0~dex,
our data are statistically insufficient to well constrain the models that could predict larger ratios than observed.

In addition, some peculiar stars are also identified in the AMBRE:Pb catalogue. Three probable Asymptotic Giant Branch
stars are found to be not enriched in lead, revealing that they still did not experience any third-dredge-up events. On the contrary, nine stars very enriched in lead are found, eight of them being known as Carbon-Enriched Metal-Poor (CEMP) stars. The last one could thus be a new CEMP candidate. Moreover, the Pb abundance of 13 \Gaia\ Benchmark Stars are also reported, adopting both the AMBRE and the literature reference values for their atmospheric parameters. 

This AMBRE:Pb catalogue is made publicly available and should be an important step forward in our understanding of the Galactic chemical evolution of very heavy neutron-captured elements and of their production rates in different types of stars.

\begin{acknowledgements}
        The authors thank L. Mashonkina for providing the NLTE corrections for the \PbI\ 368.3~nm line. We also thank T. Masseron for having shared some molecular line lists partially used within the AMBRE Project. \\

 This work has made use of data from the European Space Agency (ESA)
mission \Gaia\  (https://www.cosmos.esa.int/gaia), processed by the \Gaia\  Data Processing and Analysis Consortium (DPAC, https://www.cosmos.esa.int/web/gaia/dpac/consortium). Funding for the DPAC has been provided by national institutions, in particular the institutions participating in the \Gaia\ Multilateral Agreement. This work has also made use of the VALD database, operated at Uppsala University, the Institute of Astronomy RAS in Moscow, and the University of Vienna.
         We also used the SIMBAD database, operated at CDS, Strasbourg, France.
         Part of the calculations were performed with the high-performance computing 
         facility SIGAMM, hosted by OCA.\\

This work was supported by the Deutsche Forschungsgemeinschaft
(DFG, German Research Foundation) – Project-ID 279384907 – SFB 1245, the State of Hessen within the Research Cluster ELEMENTS (Project ID 500/10.006).\\

Finally, we are
grateful to the anonymous referee for their constructive
feedback, which has improved the clarity of the
manuscript.
\end{acknowledgements}
\bibliographystyle{aa} 
\bibliography{ref}
\end{document}